\documentclass[fleqn,usenatbib]{mnras}

\usepackage[T1]{fontenc}
\usepackage{ae,aecompl}

\usepackage{graphicx}        
\usepackage{amsmath}         
\usepackage{amssymb}         

\usepackage{booktabs}        
\usepackage{scalerel}        
\usepackage{xcolor,listings} 
\usepackage{upquote}         
\usepackage{fixmath}         
\usepackage{multirow}        

\newcommand*{\dirplot}{plot/png_300}
\newcommand*{\ext}{png}

\newcommand*{\pd}[1]{\!\times\!10^{#1}}

\newcommand*{\units}[1]{\scalebox{0.8}{(#1)}}

\newcommand*{\diff}{\mathop{}\!\mathrm{d}}
\newcommand*{\Diff}[1]{\mathop{}\!\mathrm{d^#1}}

\newcommand*{\Sim}{\sim\!}
\newcommand*{\SM}[1]{{\scaleto{\rm #1}{3.5pt}}}

\newcommand*{\var}[1]{\mbox{\footnotesize$(#1)$}}
\newcommand*{\vvar}[1]{\mbox{\footnotesize$\big(#1\big)$}}

\newcommand*{\minus}{\scalebox{0.75}[1.0]{$-$}}
\newcommand*{\plus}{\scalebox{0.75}[1.0]{$+$}}

\newcommand*{\Gband}{\mbox{$G\:\!\text{-band}$} }
\newcommand*{\Vband}{\mbox{$V\!\!\:\text{-band}$} }

\newcommand*{\VI}{V\!\textnormal{-}I_{\scaleto{\rm C}{3.5pt}}}
\newcommand*{\BPRP}{G_{\rm BP}\!-\!G_{\rm RP}}

\newcommand*{\GtrSim}{\smallrel\gtrsim}

\makeatletter
\newcommand*{\smallrel}[2][.8]{%
  \mathrel{\mathpalette{\smallrel@{#1}}{#2}}%
}
\newcommand*{\smallrel@}[3]{%
  \sbox0{$#2\vcenter{}$}%
  \dimen@=\ht0 %
  \raise\dimen@\hbox{%
    \scalebox{#1}{%
      \raise-\dimen@\hbox{$#2#3\m@th$}%
    }%
  }%
}

\definecolor{mymauve}{rgb}{0.58,0,0.82}
\definecolor{mygreen}{rgb}{0,0.6,0}

\title[Statistical detection of the tidal stream of M68]{Statistical detection of a tidal stream associated with the globular cluster M68 using \textit{Gaia} data}

\author[C. G. Palau, J. Miralda-Escud\'e]{
Carles G. Palau,$^{1}$\thanks{E-mail: cgarcia@icc.ub.edu}
Jordi Miralda-Escud\'e,$^{1,2,3}$\thanks{E-mail: miralda@icc.ub.edu}
\\
$^{1}$Institut de Ci\`encies del Cosmos, Universitat de Barcelona (UB-IEEC), Mart\'i i Franqu\`es 1, E-08028 Barcelona, Catalonia, Spain.\\
$^{2}$Instituci\'o Catalana de Recerca i Estudis Avan\c cats, E-08028 Barcelona, Catalonia, Spain.\\
$^{3}$Institute for Advanced Study, Princeton, NJ 08540, USA.
}

\date{Accepted XXX. Received YYY; in original form ZZZ}

\pubyear{2019}

\begin{document}
\label{firstpage}
\pagerange{\pageref{firstpage}--\pageref{lastpage}}
\maketitle

\begin{abstract}
A method to search for tidal streams and to fit their orbits based on maximum
likelihood is presented and applied to the \textit{Gaia} data. Tests of the method are
performed showing how a simulated stream produced by tidal stripping of a star
cluster is recovered when added to a simulation of the \textit{Gaia} catalogue. The
method can be applied to search for streams associated with known progenitors
or to do blind searches in a general catalogue. As the first example, we apply
the method to the globular cluster M68 and detect its clear tidal stream
stretching over the whole North Galactic hemisphere, and passing within 5 kpc
of the Sun. This is one of the closest tidal streams to us detected so far, and is
highly promising to provide new constraints on the Milky Way gravitational
potential, for which we present preliminary fits finding a slightly oblate dark
halo consistent with other observations. We identify the M68 tidal stream with
the previously discovered Fj\"orm stream by Ibata et al. The tidal stream is
confirmed to contain stars that are consistent with the HR-diagram of M68. We
provide a list of 115 stars that are most likely to be stream members, and
should be prime targets for follow-up spectroscopic studies.
\end{abstract}

\begin{keywords}
globular clusters: individual: M68 - Galaxy: halo - Galaxy: kinematics and dynamics - Galaxy: structure.
\end{keywords}



\section{Introduction}

 Tidal streams are a powerful method to constrain the Milky Way
potential by dynamically modelling them and obtaining fits to their
evolution as stars are released from the tidally perturbed progenitor.
Several tidal streams have been discovered containing large numbers of
stars, the majority of them from dwarf galaxies. The largest stream is
produced by the Large Magellanic Cloud \citep{1972A&A....18..224W}, and
the second is made by the tidal arms of the disrupting Sagittarius dwarf
galaxy \citep{2003ApJ...599.1082M}; these streams may be affected by
more complex physical processes including star formation in gas that is
affected by ram-pressure as it moves through the Galactic halo hot gas.
Other tidal streams do not have an obvious progenitor, such as the
Orphan stream \citep{2006ApJ...645L..37G} and the Monoceros Ring
\citep{2002ApJ...569..245N}. Tidal streams were also discovered from the
globular clusters Palomar 5 \citep{2001ApJ...548L.165O} and NGC5466
\citep{2006ApJ...637L..29B, 2006ApJ...639L..17G}. In addition, the GD-1
tidal stream, spanning about 60 deg on the sky, is believed to have
originated from a globular cluster that was finally completely disrupted
by the tidal perturbation \citep{2006ApJ...643L..17G}. Additional cases
of tidal tails have been reported which are more distant and of a lower
surface brightness than previous ones \citep[e.g.][]{2009ApJ...693.1118G}, most of them recently discovered with photometry data of Dark Energy Survey \citep{2016ApJ...820...58B, 2018ApJ...862..114S}. A list of the streams detected up until 2015 and a summary of the techniques used for finding them
is found in \citet{2016ASSL..420...87G}.

 As the number of discovered tidal streams grows and the accuracy of the
measured motions of stars within them is improved, increasingly accurate
constraints on the Milky Way potential should be derived by requiring a
match of the dynamics of all streams. Several methods have been designed
for this purpose
\citep{2011MNRAS.417..198V, 2014ApJ...795...94B, 2014ApJ...794....4P}.
One of the most notable applications of these streams that has been
discussed is the creation of density variations and gaps along tidal
streams from tidal interactions with perturbing satellites moving across
the Milky Way halo, including the predicted dark matter sub-haloes
\citep{2009ApJ...705L.223C}. The Cold Dark Matter model should in
principle make reliable predictions for the abundance and shape of any
such gaps in tidal streams, although other objects such as globular
clusters and molecular clouds may give rise to similar features 
\citep{2013ApJ...768..171C, 2018MNRAS.477.1893D}. For all these reasons,
a strong interest has developed in detecting as many streams as
possible. The publication of the \textit{Gaia} Data Release 2 (GDR2), with proper motions and parallaxes of unprecedented accuracy, is a giant step
forward in our ability to discover, analyse and model large numbers of tidal
streams in the Milky Way
\citep{2018ApJ...865...85I, 2018MNRAS.481.3442M,2018ApJ...863L..20P}.

 Globular clusters are an important class of objects that should, in
general, be accompanied by tidal streams. At each passage through the
galactic pericentre and at each crossing of the disc, the tidal
perturbation reaches a maximum and causes a peak in the rate at which
stars are tidally pulled out of the system into unbound orbits,
populating the leading and trailing tails. The streams, however, may be
of very low density and difficult to detect. Apart from tidal
truncation, globular clusters also evolve under internal dynamical
relaxation and mass segregation, which preferentially diffuses low-mass
stars into outer orbits from which they can be tidally perturbed and
incorporated into the tidal tails. This selection favouring low-mass
stars in tidal tails can make them more difficult to detect in
flux-limited surveys \citep{2000A&A...359..907L, 2018MNRAS.474.2479B}. 

 So far, tidal tails have often been found serendipitously by simply
noticing an excess of stars along bands in the sky after applying certain cuts based
on proper motions and photometry. With the massive amount of new data
coming from the \textit{Gaia} mission, there is a need to optimize and
systematize detection methods of tidal streams to ensure that they are
efficiently found. Several methods have been proposed to find them from
large samples of stars with phase-space coordinates, magnitudes, and
colours:

\begin{itemize}
\item \citet{2005MNRAS.359.1287B} try to identify tidal streams on the
energy-angular momentum plane. This method needs an axisymmetric
potential for the Galaxy and requires high-quality data because the
large background population and the observational uncertainties make
difficult the detection of a stream as a single structure projected on
only two integrals of the motion, even with \textit{Gaia} data.

\item \citet{2015ApJ...801...98S} integrate orbits in a specific
potential model of the Galaxy, and then search for streams in the
angle-action space using a clustering algorithm. This method tends to
suffer from similar problems as the previous method when using an
oversimplified potential model that scatters the real stream members
over an excessively large region of phase space.

\item \citet{2017MNRAS.469..721M} use an algorithm (Great Circle Method)
exploiting the fact that streams are confined to a plane if the Galaxy
potential is close to spherical, searching for stars with positions and
proper motions that approximately lie in a unique plane. The problems
are similar to the previous methods when the potential is not very close
to spherical and parallax errors introduce uncertainties in determining
the plane of the star orbits.

\item Finally, \citet{2018MNRAS.477.4063M} present an algorithm, called
Streamfinder, designed to find distant halo streams, where a
six-dimensional tube in phase space that follows the orbit of a star computed in a fixed potential is constructed with a width similar
to the size of the expected stream. A large number of observed stars that are
compatible with this tube, taking into account the errors of the
measurements, is then an indication of the reality of a putative
stream.
\end{itemize}

 Here we present a new method based on maximum likelihood analysis that
searches for streams in a large data set, computing orbits in a
gravitational potential that can be varied together with the kinematic
initial conditions of a stream progenitor. The method takes into account
the probability density of stars to belong to a model of the stellar
foreground and to belong to a tidal tail, and can be applied for known
candidate progenitors or for blind searches with unknown progenitor
orbits. The search for a stream is done after a pre-selection that
eliminates most of the stars in the data set that are very unlikely to
belong to any tidal stream in the phase-space region being searched,
which is used to reduce computational time. Among previous stream
detection methods, our one resembles the most the Streamfinder method of
\citet{2018MNRAS.477.4063M}, improving and generalizing the likelihood
function using a model of the phase-space density of the Milky Way for
the stellar foreground and a variable gravitational potential, and a
realistic dynamical treatment of a tidal stream taking into account the
intrinsic and observational uncertainties at the pre-selection and
best-fitting stages.

 This type of method has so far not been exhaustively applied to the
list of known globular clusters. In this paper, we develop this method
to search for tidal tails around a known progenitor, with the goal of
identifying them even though they will generally not be intuitively
visible to the eye as an excess under any projections and cuts and can
only be detected as a statistical excess above the foreground or
background stars with known observational errors. We apply the method
specifically to the globular cluster M68 (NGC4590) in this paper,
because of several characteristics that make it the best candidate for
observing any associated tidal stream. We find clear evidence of a long
stream with more than 100 member stars that promises many applications
for studies of the Galactic potential and the dynamics of tidal stream
formation.

 As this work was being completed, we became aware\footnote{
We thank Daisuke Kawata for pointing this out to us and for many
inspiring comments on our work}
that the stream we detect was already discovered in \citet{2019ApJ...872..152I}, who baptized it with the name Fj\"orm; however,
they did not associate it with the globular cluster M68. We find
incontrovertible evidence for this association, which makes this stream
even more interesting for studies of the dynamics of globular clusters
and the Milky Way potential.

 Our method is described in detail in Section \ref{sec2}, and
Section \ref{sec3} presents tests on simulations of the \textit{Gaia} catalogue,
applying them specifically to stream models of M68. In Section
\ref{sec4}, we search for a real stellar stream associated with M68
using \textit{Gaia} data and we present our conclusions in Section \ref{con}.

\section{Statistical method to detect tidal streams}\label{sec2}

  In this section we describe the method we have developed to search for
tidal streams. Our goal is to be able to detect a tidal stream even when
seen against a large number of foreground stars with similar kinematic
characteristics, and when large observational errors on the distance and
proper motions and the absence of radial velocities make it difficult to
visualize the tidal stream directly from any projection of the data. In
many cases, there may be no individual stars that can be assigned to a
tidal stream with high confidence, even though the existence of the
tidal stream itself may be highly significant from a large set of
candidate members. The method is based on the maximum likelihood
technique, although the likelihood function has to be defined in an
approximate way because of the difficulty of numerically computing
precise distributions of a tidal tail for many models, and of correctly
characterizing the foreground. The
approximations used are tested and calibrated using mocks of the \textit{Gaia} data.

  Stars in a tidal stream are stripped out from a bound object (a
globular cluster or a dwarf galaxy) by the external tidal force of the
Milky Way galaxy. Tidal stripping occurs when the mean
density of the Galaxy within the cluster orbit is comparable to the mean
cluster density, which also implies that the orbital times of the stripped
stars in the cluster are comparable to the orbital time of the cluster
around the Galaxy. Tidal stripping is therefore strongest at the closest
pericentre passages. In addition, tidal shocks occur when the cluster
crosses the disc. The escaped stars approximately follow the progenitor
orbit, with small variations of the conserved integrals of the motion.
Stars moving to lower orbital energy form a leading tail ahead of the
cluster trajectory, while stars left at a higher orbital energy form a
trailing tail behind. The shape of these tails is a first approximation
to the cluster orbit, although in detail, they are different
\citep{2010MNRAS.401..105K, 2014ApJ...795...95B, 2015MNRAS.452..301F}.

\subsection{Likelihood function and parameter estimation methodology}\label{likelihood}

  The maximum likelihood method is used to determine parameters of a
model that maximize a posterior function, given a set of observational
data. The likelihood function is the probability density of obtaining
the observed data as a function of the model parameters. The data
generally consist of $N$ independent observations of a set of $n_{\rm p}$
variables $w^\mu$ (labelled by an index $\mu=1,...,n_{\rm p}$), with a
covariance matrix for observational errors $\sigma^{\mu\nu}$. We want to
compute a probability density
$P\var{w^\mu|\theta_\kappa;\sigma^{\mu\nu}}$ in the space of the
$w^\mu$ variables, depending on $K$ model
parameters $\theta_\kappa$ (with $\kappa=1,...,K$) that are to be
estimated. The dependence on the covariance matrix is assumed to be a
convolution with Gaussian errors, in the appropriately chosen $w^\mu$
variables. In our case, $N$ is the number of
stars in a selected catalogue where we search for evidence of a tidal
tail, $w^\mu$ are the $n_{\rm p}=6$ observed coordinates of each star of
parallax, angular position, radial velocity and proper motion,
and the covariance $\sigma^{\mu\nu}$ can be different for every star.
The likelihood
function $L\var{\theta_\kappa}$ is the product of the probability
densities of all the data points:
\begin{equation}
\label{eq:lf}
L\var{\theta_\kappa} = \prod_{n=1}^{N}
 P\var{w^\mu_n|\theta_\kappa;\sigma^{\mu\nu}_n} ~.
\end{equation}
A prior probability function $p\var{\theta_\kappa}$ is assumed for
the parameters $\theta_\kappa$.

  In addition to fitting parameter values, the likelihood function can
be used to calculate a statistical confidence level to establish if a
certain hypothesis is true or false. Usually a null-hypothesis $H_0$
states that the $K$ model parameters obey a set of $T$ restricting
equations, meaning that the $K$ parameters lie in a sub-space of
dimensionality $K-T$, while the one-hypothesis $H_1$ asserts that the
parameters do not obey the $T$ restrictions and are outside of this
sub-space. Let the parameters $\hat{\theta}^0_\kappa$ and
$\hat{\theta}^1_\kappa$ be the values that maximize the posterior
function, defined as the product
$L\var{\theta_\kappa} \, p\var{\theta_\kappa}$, with the $T$
restrictions and without them, respectively. The truth of $H_0$ can be
tested using the likelihood ratio statistic $\Lambda$, defined as:
\begin{equation}\label{Lamf}
\Lambda \equiv -2 \ln \left[ \frac{
  L\var{\hat{\theta}^0_\kappa}\, p\var{\hat{\theta}^0_\kappa}}
 {L\var{\hat{\theta}^1_\kappa}\, p\var{\hat{\theta}^1_\kappa}}
\right] ~.
\end{equation}
We use Wilks' theorem \citep[see e.g.,][]{casella2002statistical}
to choose the criterion $\Lambda < k$ for favouring the null-hypothesis
$H_0$, where $k$ is computed from a conventionally chosen value of the
probability $\epsilon$ of inappropriately rejecting the null-hypothesis
when $H_0$ is actually true, using the equation
\begin{equation}
 \epsilon = \int_k^\infty\!\chi^2_T\var{z}\,\diff z =
 \frac{\Gamma(T/2, k/2)}{\Gamma(T/2)} ~,
\end{equation}
where $\chi^2_T$ is the standard $\chi^2$ distribution with $T$ degrees
of freedom. We shall use a confidence level $\epsilon=0.01$ throughout
this paper, which for one degree of freedom implies $k=6.635$.

  In general, the $N$ stars of any set in which we search for a tidal
stream contain a fraction $\tau$ of stars that belong to the tidal
stream, and a fraction $1-\tau$ that belong
to the foreground containing the general stellar population of the
Galaxy (note that what we refer to as ``foreground'' stars may actually
be foreground or background compared to the stream, and we simply mean
that they are superposed in the sky with the hypothesized stream and
belong to the set in which the stream is being searched for). The null-hypothesis is simply the restriction $\tau=0$. The probability density of the data variables for each star $n$ is
\begin{multline}\label{eq:pdsf}
 P\var{w^\mu_n |\theta_{\kappa};\sigma^{\mu\nu}_n} =
\tau \, P_S\var{w^\mu_n|\theta_s,\theta_c,\theta_\phi;\sigma^{\mu\nu}_n} \\[0.5em]
+ (1-\tau)\, P_F\var{w^\mu_n|\theta_s;\sigma^{\mu\nu}_n} ~.
\end{multline}
Here, $P_S$ is the probability density that a star belonging to the stream
has variables $w^\mu_n$, convolved with errors $\sigma^{\mu\nu}_n$,
while $P_F$ is the probability density for a star that belongs to the
foreground. We have split the $K$ model parameters into three groups:
$\theta_s$ refers to parameters of the distribution of stars in
various components of the Galaxy (disc, bulge, and halo), $\theta_c$
are the present coordinates of the orbiting object generating the tidal
stream, and $\theta_\phi$ are parameters of the Milky Way
potential. In general, the parameters $\theta_s$ also affect the
potential if the stellar components are assumed to imply a mass
component with the same distribution (i.e., if a fixed mass-to-light
ratio of the stellar population is assumed), so we include these
parameters in $P_S$ as well. The likelihood function is then given by
equations (\ref{eq:lf}) and (\ref{eq:pdsf}). The values of the parameters for which the likelihood function is maximum $\hat{\theta}_\kappa$ are computed using a Nelder-Mead Simplex algorithm, which does not require neither a smooth function nor the evaluation of its derivatives \citep[see][for details]{conn2009introduction}.
The covariance matrix for the errors of the parameter solutions $\sigma^{ij}$ are computed by calculating second derivatives of the posterior function logarithm
\begin{equation}\label{uneq}
\sigma^{ij} = \left(
 - \frac{\partial^2 \ln\! \vvar{ L\var{\theta_\kappa}\, 
 p\var{\theta_\kappa} } }
{\partial\theta_i \, \partial\theta_j} \bigg|_{\hat{\theta}_\kappa}
 \right)^{-1} ~.
\end{equation}

 The stellar phase-space density model depending on parameters $\theta_s$
is described in detail in Section \ref{MWModel}, and the calculation of
$P_F$ taking into account an observational selection approximation is
explained in Section \ref{PF}.
The computation of the probability density $P_S$ from a density model of
the stream, using the stream progenitor orbit and a model of the Milky Way 
potential, is described in Section \ref{PS}.

We have not included information about colours and magnitudes of the stars in the likelihood function since it would be necessary to model the colour-magnitude distribution of the foreground stars as seen by \textit{Gaia}. Although the inclusion of photometric data might improve the detection capability, we leave this modification for future further improvement of the statistical method. In Section \ref{sec3}, we use the phase-space data to establish the existence of a simulated tidal stream and to compute the best adjustment of the model parameters using a foreground simulated star catalogue. We use colours and magnitudes in Section \ref{sec:ids} to improve the final identification of the star candidates to belong to the stellar stream choosing those that are compatible with the HR-diagram of the progenitor cluster.


\subsection{Phase-space stellar model of the Milky Way}\label{MWModel}

 We define a simple model of the phase-space distribution of stars in the Milky Way, $f$, to enable an estimate of the probability density that a
star belongs to the Galactic foreground, $P_F$. The distribution in our
model is the sum of four components: thin disc, thick disc, bulge and stellar halo. We
write each of the four components, labelled by the index $\gamma$,
as the product of a stellar mass density function of space,
$\rho_\gamma$, and a velocity distribution function $g_\gamma$:
\begin{equation}
f\var{x_i, v_j} = \frac{1}{M} \sum_{\gamma=1}^4 \rho_\gamma\var{x_i} \, g_\gamma\var{x_i,v_j} ~,
\end{equation}
where $x_i$ and $v_j$ are three-dimensional components of position and
velocity vectors, and $M$ is the total mass of stars in the four
components. The velocity distribution $g_\gamma$ is normalized to unity
at each position $x_i$, when integrating over all the velocity space. We
make the simplified assumption that the number of stars per unit stellar
mass at a given luminosity in the \textit{Gaia} band is the same for all
components.

\subsubsection{The disc density}

The disc is constructed as the sum of two exponential profiles, for the
thin and thick disc. In cylindrical coordinates $(R,\varphi,z)$, the mass
density for each component is
\begin{equation}\label{d_profile}
\rho_\gamma\var{R,z} = \frac{\Sigma_\gamma}{2z_\gamma} 
 \exp\left( -\frac{R}{h_\gamma} -\frac{|z|}{z_\gamma} \right)~.
\end{equation}
We use the parameter values in \citet{2011MNRAS.414.2446M}, which are
listed in Table \ref{DataDisc}.

\begin{table}
\caption[]{\small{Properties of the disc density model.}}
\begin{center}
\begin{tabular}{llll}
\toprule
\multicolumn{2}{l}{\textbf{Properties}}&\textbf{Thin disc}&\textbf{Thick disc}\\
\midrule
$\Sigma_{1,2}$&\units{M$_{\odot}$ kpc$^{-2}$}&$8.17\pd{8}$&$2.1\pd{8}$\\
$h_{1,2}$&\units{kpc}&$2.9$&$3.31$\\
$z_{1,2}$&\units{kpc}&$0.3$&$0.9$\\
\midrule
$M_{1,2}$&\units{M$_{\odot}$}&$4.31\pd{10}$&$1.44\pd{10}$\\
\bottomrule
\end{tabular}
\end{center}

\begin{tabular}{l}
\textit{Note.} Ref.: \citet{2011MNRAS.414.2446M}.\\
\end{tabular}

\label{DataDisc}
\end{table}

\subsubsection{The bulge density}

  For simplicity we assume an axisymmetric bulge, even though the bulge
is a rotating bar, since our conclusions in this paper do not depend on
accurately modelling orbits in the inner part of the Galaxy. The bulge
density is a power law with slope $\alpha$ and a Gaussian truncation at
a scale length $a_1$:
\begin{equation}\label{b_profile}
\rho_3\var{s} = \rho_{0} \left(1+as\right)^{-\alpha} \exp\left(-s^2 \right) ~,
\end{equation}
where
\begin{equation}
s^2 \equiv \frac{R^2}{a_1^2} + \frac{z^2}{a_3^2} ~.
\label{ell}
\end{equation}

\noindent The bulge parameters are listed in Table \ref{DataBDHSH},
taken from \citet{2011MNRAS.414.2446M}.

\subsubsection{The stellar halo density}

The Galaxy is surrounded by a faint stellar halo made by old and metal-poor stars. We model it as an oblate ellipsoidal object with a density
profile as a function of the radial variable in equation (\ref{ell})
following a two power-law model:
\begin{equation}\label{powlaw}
\rho_{4}\var{s} = \rho_{0}\,s^{-\alpha} \, (1+s)^{\alpha-\beta} ~.
\end{equation}
We use the parameters of \citet{2014AandA...569A..13R}, listed in
Table \ref{DataBDHSH}.

\begin{table}
\caption[]{\small{Properties of the bulge, dark halo and stellar halo density models.}}
\begin{center}
\begin{tabular}{lllll}
\toprule
\multicolumn{2}{l}{\textbf{Properties}}&\textbf{Bulge}&\textbf{Stellar halo}&\textbf{Dark halo}\\
\midrule
$\rho_0$&\units{M$_{\odot}$ kpc$^{-3}$}&$9.93\pd{10}$&$2.66\pd{3}$&$\rho_{0 \rm dh}$\\
$a_1$&\units{kpc}&$2.1$&$2.1$&$a_{1 \rm dh}$\\
$a_3$&\units{kpc}&$1.05$&$1.68$&$a_{3 \rm dh}$\\
$a$&&$28$&$-$&$-$\\
$\alpha$&&$1.8$&$1$&$1$\\
$\beta$&&$-$&$3.8$&$\beta_{\rm dh}$\\
\midrule
$M$&\units{M$_{\odot}$}&$8.96\pd{9}$&$1.72\pd{5}$&$-$\\
\bottomrule
\end{tabular}
\end{center}

\begin{tabular}{l}
\textit{Note.} Bulge Ref.: \citet{2011MNRAS.414.2446M}.\\
Stellar Halo Ref.: \citet{2014AandA...569A..13R}.
\end{tabular}

\label{DataBDHSH}
\end{table}

\subsubsection{The velocity distribution model}

  For all the stellar components, we assume that the velocity
distribution function is a Gaussian with principal axes oriented along
spherical coordinates,
\begin{multline}
g_\gamma\var{v_r,v_\theta,v_\phi} =
 \frac{1}{(2\pi)^{3/2}\,\sigma_{\gamma r}\sigma_{\gamma\theta}
  \sigma_{\gamma\phi} }\, \\
 \exp\left( -\frac{v_r^2}{2\sigma_{\gamma r}^2} -
  \frac{v_\theta^2}{2\sigma_{\gamma\theta}^2} -
 \frac{(v_\phi-v_{\rm ad})^2}{2\sigma_{\gamma\phi}^2} \right) ~,
\end{multline}
where the index $\gamma$ labels the four stellar components.
The three-velocity dispersion eigenvalues are determined observationally.
We use the values for each component from \citet{2012AandA...543A.100R},
assuming the average of all stellar populations for the thin disc. The values
are listed in Table \ref{DataVelocity}.

\begin{table}
\caption[]{\small{Velocity dispersions of the bulge, dark halo and stellar halo density models and asymmetric drift.}}
\begin{center}
\begin{tabular}{lllll}
\toprule
&$\sigma_{r}$&$\sigma_{\theta}$&$\sigma_{\phi}$&$v_{\rm ad}$\\
&\units{km s$^{-1}$}&\units{km s$^{-1}$}&\units{km s$^{-1}$}&\units{km s$^{-1}$}\\
\midrule
\textbf{Thin disc}&31&12.6&20&-229.4\\
\textbf{Thick disc}&67&42&51&-185\\
\textbf{Bulge}&113&100&115&-159\\
\textbf{Stellar halo}&131&85&106&-12\\
\bottomrule
\end{tabular}
\end{center}

\begin{tabular}{l}
\textit{Note.} Ref.: \citet{2012AandA...543A.100R}.\\
\end{tabular}

\label{DataVelocity}
\end{table}


\subsection{Potential model of the Milky Way}\label{secpot}

 We assume the dark matter halo density profile also follows equation
(\ref{powlaw}). For the case $\alpha=1$ and $\beta=3$, this density
profile reduces to the NFW profile \citep{1996ApJ...462..563N}, which
fits the dark matter halo profiles obtained in cosmological
simulations. For simplicity we also assume an axisymmetric shape for
the dark halo, even though in general it can be triaxial. In the
models in this paper we fix $\alpha=1$ and we leave the other
parameters listed in Table \ref{DataBDHSH} to be free when fitting the
model. In practice this gives enough freedom to our halo profile to
model the dynamics of tidal tails we examine here.

 Two baryonic components are added to this dark halo: the disc and the
bulge. Their potentials are computed following the previous stellar
density profiles in equations (\ref{d_profile}) and (\ref{b_profile}). The
stellar halo and gas components are neglected because of their expected small mass compared to
the dark halo.


\subsection[The probability function PF]
{The probability function $\boldsymbol{P_\SM{F}}$}\label{PF}

 In general, stellar surveys provide measurements of phase-space
coordinates of stars in the form of parallaxes $\pi$, angular positions $\delta$ and $\alpha$,
radial velocity $v_r$, and proper motions $\mu_\delta$ and $\mu_\alpha$. These are our six-dimensional
variables for each star,
$w^\mu= (\pi,\delta,\alpha,v_r,\mu_\delta,\mu_\alpha)$, where the
proper motions are $\mu_\delta = \diff\delta/\diff t$ and
$\mu_\alpha = \diff\alpha/\diff t$. Note that the physical proper motion
component in right ascension is
$\mu_{\alpha *} = \mu_\alpha \cos\delta$. The \textit{Gaia} mission is
at present providing the largest star survey. We designate as $w^\mu_o$
the observed value of each variable, and their observational errors are
characterized by a covariance matrix $\sigma^{\mu\nu}$. The values of
the true variables $w^\mu$ are assumed to follow Gaussian distributions
that we write as $G(w^\mu-w^\mu_o|\sigma^{\mu\nu})$.

 To calculate the probability density of the
observed coordinates for foreground stars, we need to take into account
the flux-limited survey selection. Let $\psi_{\rm s}\var{L}$ be the cumulative
luminosity function of stars with a luminosity in the observed
photometric band greater than $L$. Neglecting the effect of dust
absorption (which generally has small variations for the halo stars we
are interested in),
the density of stars included in the survey as a function of the
heliocentric distance $r_{\rm h}$ is proportional to
$\psi_{\rm s}\var{L_1 r_{\rm h}^2/r_1^2}$, where $L_1$ is the luminosity corresponding
to the survey flux threshold at a distance $r_1$. We can take the
normalizing distance $r_1$ to be 1 parsec, and then the parallax
expressed in arc seconds is $\pi=r_1/r_{\rm h}$.
We will assume here as a simple model that $\psi_{\rm s}\var{L}
\propto L^{-1}$ in the range of interest, meaning that there is a
roughly constant luminosity coming from stars in any range of
$\diff\log \!L$. This is roughly correct for stellar populations lying
between the main-sequence turn-off and the tip of the red giant branch.
A detailed modelling of the foreground to obtain an
accurate estimate of the likelihood function would clearly require
a more careful evaluation of the stellar luminosity function, but given
all the uncertainties in our modelling (i.e., the velocity distribution,
dust absorption, etc.) we decide to use this very approximate and simple
approach in this work.

  The foreground probability density can then be written as
\begin{multline}\label{Pmw}
 P_\SM{F}\var{w^\mu|\theta_s,\sigma^{\mu\nu}} =
\frac{1}{C}\, \int \Diff{6} w \, f\var{w^\mu|\theta_{s}}\, \\
 \frac{r_1^5 \cos^2\!\delta}{\pi^6} \; \psi_{\rm s}\var{L_1/\pi^2} \; G\var{w^\mu-w^\mu_o |\sigma^{\mu\nu}} \, ,
\end{multline}
where the Jacobian of the transformation from the cartesian $(x_i,v_j)$
variables of phase space to the $w^\mu$ coordinates has a space part
$(r_1^3\cos\delta)/\pi^4$, and a velocity part $(r_1^2\cos\delta)/\pi^2$.
The constant $C$ renormalizes the probability density in the observed
variables after the flux-threshold selection is included, and is
\begin{equation}\label{Pmr}
 C = \int \Diff{6} w \, 
  \frac{r_1^5 \cos^2\!\delta}{\pi^6} \; \psi_{\rm s}\var{L_1/\pi^2} \;
    f\var{w^\mu|\theta_{s}}\, ~.
\end{equation}
For our choice $\psi_{\rm s}\var{L}\propto L^{-1}$, the function $\psi_{\rm s}$ is
simply replaced by $\pi^2$ in equations (\ref{Pmw}) and (\ref{Pmr}).

We now assume that the observational errors are dominated by the
parallax error $\varepsilon_{\pi}$ and the radial velocity $\varepsilon_{v_r}$, neglecting errors in
proper motion compared to the intrinsic velocity dispersion of stars.
Errors in the angular positions are always negligibly small.
The integral yielding our foreground probability density is then
\begin{multline} \label{pfeq}
 P_\SM{F}\var{w^\mu|\theta_{s}} =
 \frac{r_1^5 \cos^2\delta}{C}
 \int_0^\infty \frac{\diff\pi}{\pi^4} \! \int_{-\infty}^\infty \diff v_r \; \\
 G\var{\pi-\pi_o|\varepsilon_{\pi}^2} \;
 G\var{v_r-v_{ro}|\varepsilon_{v_r}^2} \;
 f\var{x_i,v_j|\theta_{s}} ~.
\end{multline}
In the absence of any radial velocity measurement, we can
simply use a very large value of $\varepsilon_{v_r}$, or redefine the
probability $P_\SM{F}$ by integrating over all radial
velocities. Our assumption of small proper motion errors is not always
valid, and in this case, an improved estimate needs to integrate over
the full three-dimensional velocity distribution.

  The integral in equation (\ref{pfeq}) is computed numerically,
converting the observable coordinates $w^\mu$ to heliocentric
cartesian coordinates and Galactocentric ones as described
in Appendix \ref{App2} to evaluate $f\var{w^\mu|\theta_{s}}$.


\subsection[The probability function PS]
{The probability function $\boldsymbol{P_\SM{S}}$}\label{PS}

The most accurate way to model tidal streams is through direct $N$-body
simulations. However, these require the introduction of a softening radius
when particles are a random representation of collisionless matter instead
of real stars. Including a sufficient number of particles to model the
evolving gravitational potential of the globular cluster and its tidal tail
would increase the computational time by more than a factor $\Sim 10$ even with
the use of a tree-code or other techniques. In addition, we should include the perturbation on the
Milky Way potential by the globular cluster because it would be of similar importance, thus increasing the computational requirements of the model.
In practice, we need to compute the trajectories of thousands of stars
in a tidal tail, and to repeat the calculation for hundreds of models to
minimize the likelihood function and to obtain a fit. To make this
computationally feasible, we neglect the self-gravity of the stream and
compute test particle trajectories in the fixed Milky Way potential plus the
fixed potential of the satellite system orbiting around the Milky Way,
neglecting any dynamical perturbation on the satellite and on the Milky Way.

 Based on these simplifications, a wide range of studies are based on
releasing test particles near the Lagrange points, with a random offset
in position and velocities following Gaussian distributions \citep[see e.g.,][]{2012MNRAS.423.2845L}. This method reasonably reproduces
$N$-body simulations with a large reduction in computational time, and has
been studied in detail in \citet{2008MNRAS.387.1248K} and in
\citet{2012MNRAS.420.2700K}. The potential of the progenitor is
sometimes not taken into account in these simulations,
even though its effects can be significant, as was shown for example in
\citet{2014MNRAS.445.3788G}.

 In this work, we use neither $N$-body simulations because of their
impractical computational demand, nor the above-mentioned algorithms
because we want to simulate the kinematic structure of streams in a
general orbit and general potential for the Milky Way. We simulate
trajectories of test particles including an unperturbed potential for
the progenitor system, following these steps:
\begin{enumerate}
\item Compute backwards in time the orbit of the progenitor globular
cluster (or other bound satellite) from a reasonably well-known present
position and velocity.
\item Spread out stars around the globular cluster using a model derived
from its internal stellar phase-space distribution function.
\item Compute forwards in time the orbits of the stars within the
potential of the Galaxy, including the moving potential of the
globular cluster with its mass fixed.
\item Use the stars that have escaped from the progenitor to create a model of the
phase-space density of the stellar stream.
\end{enumerate}

\subsubsection{Simulation of the tidal stream}\label{simuTS}

 We generally refer to the stellar system being tidally stripped as a
globular cluster, although tidal streams can of course be formed by any stellar systems orbiting around our Galaxy. We assume the globular
cluster is initially in dynamic equilibrium, spherical, and with an
isotropic velocity dispersion, and adopt the Plummer Model for its
internal structure, with two parameters: the total cluster mass $M_{\rm gc}$
and a core radius $a_{\rm gc}$. The density profile as a function of the
distance to the cluster centre, $r_{\rm gc}$, is
\begin{equation}\label{plummerd}
\rho\var{r_{\rm gc}} = \frac{3M_{\rm gc}}{4\pi a_{\rm gc}^3}
  \left( 1+ \frac{r_{\rm gc}^2}{a_{\rm gc}^2} \right)^{-5/2} ~.
\end{equation}

 The velocity distribution at any radius can be expressed in terms of
the modulus of the escape velocity,
\begin{equation}\label{escv}
v_{\rm esc} = \sqrt{2\,\varPhi\var{r_{\rm gc}}} ~,
\end{equation}
where the gravitational potential is
\begin{equation}\label{ppot}
 \varPhi_{\rm gc}\var{r_{\rm gc}} = \frac{GM_{\rm gc}}{\sqrt{ r_{\rm gc}^2 + a_{\rm gc}^2 } } ~.
\end{equation}
The probability distribution of the modulus of the velocity $v$ is
\begin{equation}\label{plummerv}
g_{\rm gc}\var{v} = \frac{512}{7\pi} \,\frac{v^2}{v_{\rm esc}^2}
 \left( 1- \frac{v^2}{v_{\rm esc}^2} \right)^{7/2} ~.
\end{equation}

 We first compute orbits for a large number of test particles with random
initial conditions, obtained by generating an initial radius $r_{\rm gc}$
according to the density profile of equation (\ref{plummerd}), a velocity modulus
according to equation (\ref{plummerv}), and two random angles for the position
from the cluster centre and for the velocity vector. The orbits of the test
particles are computed in the combined potential obtained by adding that
of the Milky Way (as described above) and that of the cluster Plummer
model in equation (\ref{ppot}). The orbit of the globular cluster in the
Milky Way potential is computed first, also as a test particle and stored. The test particles
are computed next by considering the cluster potential to be fixed in
shape (neglecting the changes in the mass distribution caused by the
tidal perturbation) and moving along the computed orbit.

 A technical problem appears because most of the particles generated in
this way have orbits close to the cluster core, which is typically much
smaller than the tidal radius of the cluster, leaving only a very small
fraction of stars that can escape. Moreover, integrating the trajectory
of test particles near the cluster core is computationally expensive
because orbital periods in the cluster core are usually much shorter
than the orbital period around the Milky Way, so many short time-steps
are required. To avoid this problem, we restrict the generated test
particles to a subset representing cluster stars that are more likely to
escape than the majority, while making sure that particles that are not
selected would very rarely escape and not significantly contribute to
stars in the simulated tidal tail.

 Although it is not possible to determine analytically if a star in a
model cluster with any orbit is able to escape, we can obtain an
approximate restricted region of phase space where most escaping stars
should be located. For this purpose, we consider the restricted circular
3-body problem of a cluster in a circular orbit, where the combined
potential is time independent in the rotating frame following the
cluster motion. Considering a characteristic cluster orbital radius
$R_{\rm c}$, the distance between the centre of the cluster and the first
Lagrange point \citep[see][]{2011MNRAS.418..759R} is approximately given
by the tidal radius:
\begin{equation}\label{eqrt}
r_{\rm t} \equiv R_{\rm c}\left(\frac{M_{\rm gc}}{3M}\right)^{\frac{1}{3}} ~.
\end{equation}
With respect to the rotating reference frame where the two bodies are at
rest, the movement of a test particle can be described adding an extra
term to the potential needed to account for the centrifugal force giving
the following effective potential in spherical Galactocentric coordinates:
\begin{equation}
\varPhi_{\rm eff}\var{r} = \varPhi_\SM{MW}\var{r} + \varPhi_{\rm gc}\var{r} + \frac{1}{2}\frac{GM}{R_{\rm c}^3}r^2 ~,
\end{equation}
where $G$ is the gravitational constant and $\varPhi_\SM{MW}$ is the potential of the Milky Way. Using the theorem of conservation of energy, we define a limiting velocity taking a fixed radius $R_{\rm c}$:
\begin{equation}\label{vlim}
v_{\rm lim} \equiv \sqrt{2\,\varPhi_{\rm eff}\var{r}-2\,\varPhi_{\rm eff}\var{R_{\rm c}-r_{\rm t}}}
\end{equation}
For the simulation of the density of a tidal stream we use a sample generated via a Monte Carlo method taking only the stars with initial velocity such that $v>v_{\rm lim}$ and $r>r_{\rm t}$. The stars of the globular cluster are considered test particles and their orbits are integrated using a Runge-Kutta scheme from the past to the current position within the potential of the Milky Way and within the potential of the globular cluster keeping its mass constant.

\subsubsection{Density model of the tidal stream}\label{DM_TS}

 Once the orbits of the stars are computed, we construct a probability
density function of the tidal stream in the space of the directly
observed variables $(\pi,\alpha,\delta,v_r,\mu_\alpha,\mu_\delta)$,
designated as
$p_\SM{S}\var{w^\mu| \theta_{s}, \theta_{c}, \theta_{\phi}}$.
We use a Kernel Density Estimation method, with a Gaussian as a kernel.
If $N_{\rm e}$ is the total number of escaped stars at the present time, the
stream probability density is
\begin{equation}\label{pstream}
p_\SM{S}\var{w^\mu|\theta_{s}, \theta_{c}, \theta_{\phi}}
 = \frac{1}{N_{\rm e}} \sum_{i=1}^{N_{\rm e}} G\var{w^\mu-w^\mu_{{\rm c}i},\Xi_i^{\mu\nu}} ~.
\end{equation}
The centre of each Gaussian distribution, $w^\mu_{{\rm c}i}$, is the current
position of the simulated stream star, and the covariance matrix
$\Xi_i^{\mu\nu}$ is computed from the distribution of positions of
neighbouring stream stars, using weighting factors $c_{ij}$ that average
over neighbours out to some characteristic kernel size:
\begin{equation}
\Xi_i^{\mu\nu} = \left( \sum_{j=1}^{N_{\rm e}} c_{ij} \right)^{-1} \,\,
 \sum_{j=1}^{N_{\rm e}} c_{ij} \; (w^\mu_{{\rm c}j} - w^\mu_{{\rm c}i})
 \; (w^\nu_{{\rm c}j} - w^\nu_{{\rm c}i}) ~.
\end{equation}
The weights are defined depending on the distance between every pair of
stream stars,
\begin{equation}
c_{ij} = \left( d_0+d_{ij} \right)^{-9/2} ~; \qquad\quad
 d_{ij}^2 = \sum_{l=1}^{3} (x^l_{{\rm c}j}-x^l_{{\rm c}i})^2 ~,
\end{equation}
where $x^l_{{\rm c}i}$ are the Cartesian space coordinates of each stream star
at the present time in the simulation. We have tested that a value
$d_0=250\, {\rm pc}$ and the exponent $9/2$ in the previous equation
gives a reasonable reproduction of the shape and density profile of the
stellar stream, and we use these values in this work. These
quantities can, however, be varied to optimize any specific application of the
method. 

 Note that although we use the physical distances $d_{ij}$ to compute
the kernel weights, the stream structure is modelled in the space of
directly observed coordinates $w^\mu$. The probability density
$p_\SM{S}$ that we model as the sum of Gaussians in equation (\ref{pstream}) is therefore the product of a phase-space density times
the Jacobian to transform from phase-space Cartesian coordinates to the
$w^\mu$ variables of the observations.

\subsubsection[The probability function PS]{The probability function $P_\SM{S}$}\label{PSf}

 We can now write the probability density that a star belonging to the
stream and following the distribution as modelled in Section \ref{DM_TS}
is observed to have the variables $w^\mu_o$:
\begin{multline}
P_\SM{S}\var{w^\mu|\theta_{s}, \theta_{c}, \theta_{\phi};
\sigma^{\mu\nu}} =
 \frac{1}{C}\int \! G\var{w^\mu-w^\mu_{o}|\sigma^{\mu\nu}} \\
 \psi_{\rm s}\var{L_1/ \pi^2} \;
 p_\SM{S}\var{w^\mu|\theta_{s}, \theta_{c}, \theta_{\phi}}
 \, \Diff{6} w ~.
\end{multline}

\noindent We assume that the dispersion of the stream is much smaller
than the parallax observational error of a star, i.e.,
$\Xi_i^{\pi\pi} \ll \sigma^{\pi\pi}$, to approximate
$\psi_{\rm s}\var{\pi} \simeq \psi_{\rm s}\var{\pi_{{\rm c}i}}$. In this case the integral
in equation (\ref{integral_eq}) is easily performed because the
convolution of two Gaussians is a Gaussian with the sum of the
dispersions. Using this result, the probability of a star to have the observed position $w^\mu$
with observational errors $\sigma^{\mu\nu}$, assuming it is a member
of the stream, is
\begin{multline}\label{integral_eq}
P_\SM{S}\var{w^\mu|\theta_{s}, \theta_{c}, \theta_{\phi};
\sigma^{\mu\mu}} = \frac{1}{C} \, \sum_{i=1}^{N_{\rm e}}
 \psi_{\rm s}\var{L_1 / \pi_{{\rm c}i}^2} \\ G\var{w^\mu-w^\mu_{{\rm c}i}
|\sigma^{\mu\nu}\!+\Xi_i^{\mu\nu}}
\end{multline}
where the normalization constant is:
\begin{equation}
C = \sum_{i=1}^{N_{\rm e}} \psi_{\rm s}\var{L_1/\pi_{{\rm c}i}^2} ~.
\end{equation}

\subsection{Definition of the prior function}

 We use the prior function given by the direct observational
measurements of the distance, proper motions and radial velocity of M68,
with their quoted errors, as given in Table \ref{M68_DH} below. These
four globular cluster orbit parameters, labelled as
$\theta_c$ with $c=1,\ldots,4$, follow Gaussian distributions:
\begin{equation}\label{prif}
p\var{\theta_c} = \prod_{c=1}^4 G\vvar{\theta_c-\theta_{c{\rm o}}|\sigma^2_c} ~,
\end{equation}
with observed values $\theta_{c{\rm o}}$ and uncertainties $\sigma_c^2$ listed
in Table \ref{M68_DH}. We assume a flat prior for the four gravitational
potential parameters of the dark halo in the same table.

 Given the values $\hat{\theta}_c$ that maximize
the posterior function in equation (\ref{Lamf}), the deviation with
respect to the observed measurements can be quantified by
\begin{equation}\label{eqqu}
Q \equiv 2\ln\!\left(
 \frac{p\var{\hat{\theta}^0_c}}{p\var{\hat{\theta}^1_c}} \right) =
 \sum_{c=1}^4 \frac{(\hat{\theta}^1_c-\theta_{c{\rm o}})^2}{\sigma^2_c} ~.
\end{equation}
A value of $Q$ substantially larger than the number of parameters 4
would mean that the fit to a detected stream is not consistent with the
measured distance and velocities of the globular cluster progenitor,
indicating perhaps an inadequate potential parameterization or an
underestimation of the errors.


\section{Validation of the statistical method}\label{sec3}

  In this section we simulate the tidal stream of a globular cluster and the
way it can be observed in the \textit{Gaia} survey to test if our algorithm is
able to detect the stream against a simulated foreground and recover
some of the parameters of the progenitor orbit and the gravitational
potential in which the stream moves. 

\subsection{Description of the simulated \textit{Gaia} catalogue}

 We start describing a simulation of the entire \textit{Gaia} catalogue that we
then use to generate a realistic set of foreground stars that
a tidal tail would be observed against.

 The \textit{Gaia} Universe Model Snapshot (GUMS) is a simulated catalogue of the
full sky \textit{Gaia} survey for stellar sources, which is useful for testing
many types of statistical studies. Based on Besan\c{c}on Galaxy model
\citep{2003A&A...409..523R}, the simulation includes parallaxes,
kinematics, apparent magnitudes, and spectral characteristics for
$\Sim 1$ billion objects with \Gband magnitude $G\leqslant20$ mag.
A complete statistical analysis of the 10th version of this catalogue is
found in \citet{2012AandA...543A.100R}.

 The \textit{Gaia} Object Generator (GOG) is a simulation of the contents of the
final \textit{Gaia} catalogue based on the sources in the GUMS, and provides the
\textit{Gaia} expected measurements of astrometric, photometric, and spectroscopic
parameters with observational errors based on the \textit{Gaia} performance
models\footnote{\url{https://www.cosmos.esa.int/web/gaia/science-performance}}.
These estimated observational errors depend on instrument capabilities,
stellar properties, and the number of observations. A statistical
analysis of this catalogue was presented in \citet{2014A&A...566A.119L}.
We use the 18th version of this
simulation\footnote{\url{https://wwwhip.obspm.fr/gaiasimu/}} (GOG18) in
this work.

 The GOG18 simulates the end-of-mission \textit{Gaia} catalogue, so the predicted
errors are smaller than in the current available version (GDR2). We
correct the GOG18 uncertainties by scaling them to make them applicable
to the GDR2 catalogue. Given the simulated observational error
$\varepsilon_\mu$ of the phase-space coordinates, we obtain the scaled
uncertainties $\lambda_\mu \varepsilon_\mu$, where the scale factor is
computed to match the resulting distribution of simulated errors to the
real distribution of GDR2 errors. For the variables
$(\pi,\alpha,\delta,v_r,\mu_\alpha,\mu_\delta)$, we find the required
six scale factors to be
\begin{equation}\label{eqlam}
\lambda_\mu = (1.4,1.4,1.4,0.4,4.5,4.5) ~.
\end{equation}

  Radial velocities are most frequently not available, and in this work we
simply set the error to a large enough value to make our results
insensitive to it. We choose $\varepsilon_{v_r} = 10^3$ km s$^{-1}$ and then,
we set the observed value of the heliocentric radial velocity to zero
for all stars (this value is irrelevant when the radial velocity error is
large enough); we have tested that this value of $\varepsilon_{v_r}$ is large
enough that our results are not affected.

\subsection{The M68 globular cluster and its tidal stream}\label{M68_gc}

 Among several globular clusters we have considered as targets for a search for
associated tidal tails, M68 (NGC4590) stands out because of its proper motions and distance
are measured with good precision, it is projected on to the halo, its relative proximity to us, and a predicted orbit that brings
it close to us and keeps it far from the inner region of the Galaxy. All
these properties make any putative tidal tail easier to find: a long
orbital period far from the Galactic centre means that the tidal tail
has not been strongly broadened and dispersed by phase mixing, and a
small distance to at least part of the tidal tail allows us to discover
member stars of relatively low luminosity, especially if the foreground
is far from the plane of the Galactic disc. After going through the list
of known globular clusters, we selected M68 to be the most promising
candidate for finding an associated tidal tail for these reasons.

 The parameters for the computed orbit of the globular cluster M68 in
our model of the Galactic potential are listed in Table \ref{M68_DH},
including its pericentre, apocentre, and angular momentum component
perpendicular to the Galactic plane. As with other globular clusters,
the proper motion measured by \textit{Gaia} has dramatically improved the
precision of the predicted orbit, which has been computed for several
potential models \citep{2018AA...616A..12G}. In all the models the
obtained orbit has an apocentre $\Sim 30$ kpc, pericentre $\Sim 7$ kpc,
and a radial period of $\Sim 400$ Myr, producing an elongated stream with
little phase mixing and breadth. The globular cluster is approaching
us and will come within $\Sim 5$ kpc of the present position of the Sun
in about 30 Myr, at the time when it is near its pericentre,
flying almost vertically above our present position. This implies that
any tidal tail should have a leading arm along this future part of the
orbit, extending over the northern Galactic hemisphere, with the
trailing arm being more difficult to see because of its position closer
to the Galactic plane and further from us.

\begin{table}
\caption[]{\small{Mass, core radius, present position, and velocity of
the M68 (NGC4590) globular cluster; parameters of the dark halo mass
density used in our simulation of the tidal stream, and computed
properties of the orbit integrated for 10 Gyr.}}
\begin{center}
\begin{tabular}{lllc}
\toprule
\multicolumn{3}{l}{\textbf{Properties M68}}&\textbf{Ref.}\\
\midrule
$M_{\rm gc}$&\units{M$_{\odot}$}&$(5.7\pm2.7)\pd{4}$&[1]\\
$a_{\rm gc}$&\units{pc}&$6.4\pm2$&[1]\\
\midrule
$r_{\rm h}$&\units{kpc}&$10.3\pm0.24$&[2]\\
$\delta$&\units{deg}&$-26.7454$&[3]\\
$\alpha$&\units{deg}&$189.8651$&[3]\\
$v_r$&\units{km s$^{-1}$}&$-94.7\pm0.2$&[2]\\
$\mu_\delta$&\units{mas yr$^{-1}$}&$1.7916\pm0.0039$&[3]\\
$\mu_\alpha$&\units{mas yr$^{-1}$}&$-3.0951\pm0.0056$&[3]\\
\midrule
\multicolumn{3}{l}{\textbf{Properties dark halo}}&\\
\midrule
$\rho_{0 \rm dh}$&\units{M$_{\odot}$ kpc$^{-3}$}&$8\pd{6}$&\\
$a_{1 \rm dh}$&\units{kpc}&$20.2$&\\
$a_{3 \rm dh}$&\units{kpc}&$16.16$&\\
$\beta_{\rm dh}$&&$3.1$&\\
\midrule
$M_{200}$&\units{M$_{\odot}$}&$8.32\pd{11}$\\
$q$&&$0.8$\\
$q_{\varPhi}$&&$0.91$\\
\midrule
\multicolumn{3}{l}{\textbf{Properties orbit}}&\\
\midrule
$r_{\rm peri}$&\units{kpc}&$6.87$&\\
$r_{\rm apo}$&\units{kpc}&$35.14$&\\
$L_{z}$&\units{km s$^{-1}$ kpc}&$-2397.48$&\\
\bottomrule
\end{tabular}
\end{center}

\begin{tabular}{l}
\textit{Note.} \text{[1]}: \citet{2010MNRAS.406.2732L}.\\
\text{[2]}: \citet{1996AJ....112.1487H} (2010 Edition\footnotemark).\\
\phantom{\text{[2]}:} Relative heliocentric distance error: $\Sim 2.3$ per cent.\\ 
\text{[3]}: \citet{2018AA...616A..12G}.\\
\end{tabular}

\label{M68_DH}
\end{table}

\footnotetext{\url{http://physwww.mcmaster.ca/~harris/mwgc.dat}}

\subsubsection{Simulation of the M68 tidal stream}\label{M68_simu}

 To simulate the tidal tail produced by M68, we use the Milky Way dark
halo model described by equation (\ref{powlaw}) with parameters listed
in Table \ref{M68_DH}, corresponding to a total mass
$M_{200} \sim 8\pd{11}\, $M$_\odot$ and an axial ratio
$q \equiv a_{3 \rm dh}/a_{1 \rm dh} = 0.8$, and an implied potential flattening
$q_{\varPhi} = 0.91$ at the position of M68. The baryonic components of
the mass distribution are added as described in Section \ref{secpot}.

\begin{figure*}
\begin{tabular}{c}
\includegraphics[]{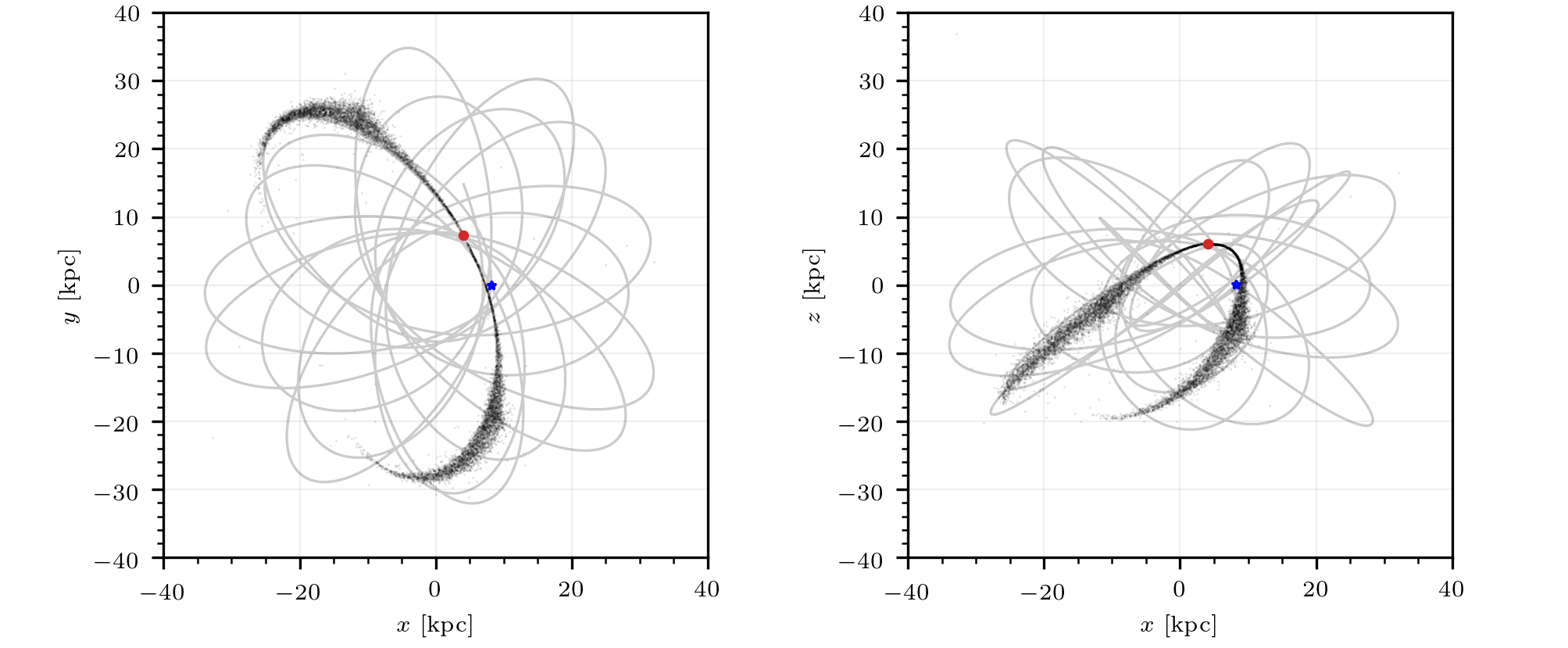}\\
\includegraphics[]{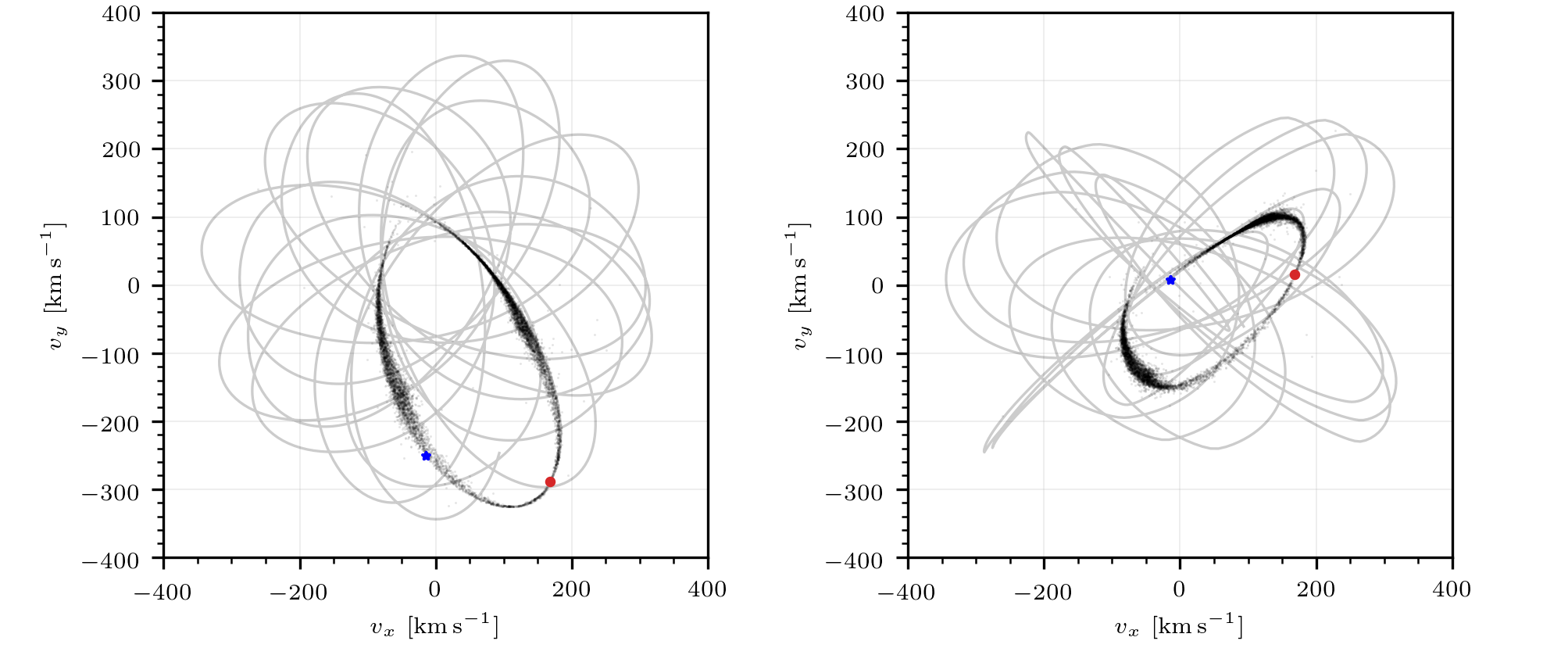}
\end{tabular}
\caption{Simulated orbit of M68 from 10 Gyr ago to the present time in
the Galactic disc plane ($x,y$) and in the perpendicular plane ($x,z$),
and the simulated tidal stream (sample of $10^4$ stars). The blue star
is the present solar position and the red dot is the position of M68.}
\label{stream_M68}
\end{figure*}

 The tidal stream is simulated following the steps described in Section \ref{PS}, without applying the cuts in the simulated test
particles described at the end of Section \ref{simuTS}. These cuts are
only applied later when the tidal tail needs to be computed for many
models and the required computational time needs to be reduced. The
orbit of M68, with properties listed in Table \ref{M68_DH}, is shown as
a grey curve in Fig. \ref{stream_M68} in two projections in space and
two projections in velocity space. The blue star indicates the present
position of the Sun and the red circle is the present position of M68.
We calculate the orbits of a total of $10^6$ test stars, out of which
$N_{\rm e} =$ 68 839 escape the potential of the globular cluster and form the
tidal tail, shown in Fig. \ref{stream_M68} as black dots.
The simulation is run over 10 Gyr, first retracing the orbit of the
globular cluster backwards in time and then calculating the orbits of
the test stars starting with the initial conditions of the Plummer model
10 Gyr ago. At the beginning of the simulation, many stars are
released because no radial cut-off is imposed on the assumed Plummer
model in the initial conditions. The first pericentre passage and disc
crossings also release many stars, but the rate of escape is reduced
later once the globular cluster has already been stripped. For this
reason, more stars are produced at the edges of the tidal tails, which
also broaden with increasing distance from the cluster. This large
number of stars released in the first orbits is a feature that obviously
depends on the history of the cluster and the Milky Way, which our model
does realistically predict. The part of the tidal tail closer to the
cluster, released at later times, should be more realistic because it is
produced after the cluster has already reached a steady rate of escaping
stars.

\begin{figure}
\begin{tabular}{c}
\includegraphics[width=\columnwidth]{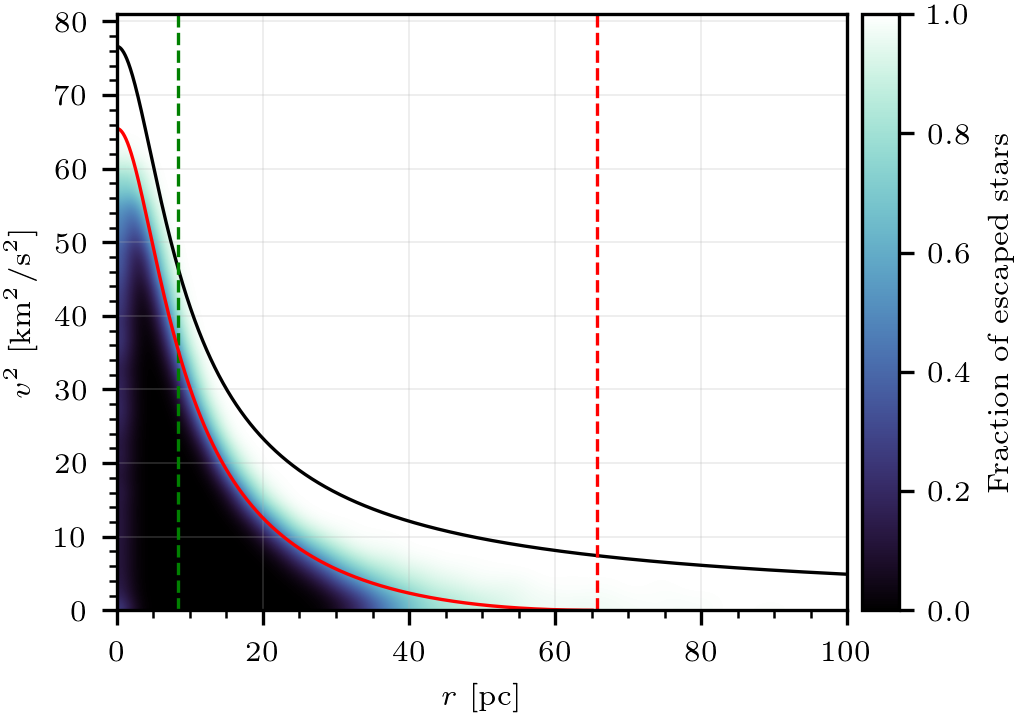}\\
\end{tabular}
\caption{Fraction of escaped stars after 10 Gyr. The green dashed line
is the half mass radius $r_{\rm h}$ and the dashed red line the tidal radius
for $R_{\rm c}=21$ kpc. The black line is the escape velocity $v_{\rm esc}$
and the red line is the limiting velocity $v_{\rm lim}$.}
\label{dist}
\end{figure}

  Among the cluster stars generated according to the initial conditions
of equations (\ref{plummerd}) and (\ref{plummerv}), the fraction of escaped
stars at the end of the 10 Gyr simulation is shown in Fig. \ref{dist}.
This figure shows that our criterion to select only stars that are
likely to escape ($v > v_{\rm lim}$, shown as the red line), used in
later simulations, is adequate to include most of the stars in the
cluster that will actually be released in the tidal tail.

\subsubsection{Simulation of the \textit{Gaia} observational uncertainties for
stream stars}

 We now include realistic observational uncertainties in the simulated
stream stars according to the expected measurement errors in the \textit{Gaia}
mission. We apply this to the total number $N_{\rm e} =$ \mbox{68 839} of escaped
stars in our simulation to simulate the observed tidal tail. Naturally,
the total number of stars in the tidal tail is highly uncertain, but
our estimate of the number of available stars can be a reasonable one
to the extent that the total number of globular cluster stars we have
simulated is comparable to the expected number of stars in M68, and
that the rate of escaping stars is also reasonable.

 The \textit{Gaia} observational uncertainties $\sigma_\mu$ for the stream stars
have been simulated using the Python toolkit
PyGaia\footnote{\url{https://pypi.org/project/PyGaia/}}, which
implements the \textit{Gaia} performance models. Observational errors depend on
the \textit{Gaia} \Gband magnitude, the Johnson-Cousins \Vband
magnitude, the colour index $\VI$, and the spectral type of each star.
These magnitudes have been simulated as follows:

\begin{enumerate}
\itemsep1.0em
\item Assign randomly to each star a \Gband absolute
magnitude and a $\BPRP$ colour index following the HR-diagram of M68
that we show in Fig. \ref{HR_M68}, which we have generated using
2929 GDR2 stars (see Appendix \ref{App4}).

\item Compute the colour index $\VI$ and the \Vband magnitude from the
following approximations \citep{2010A&A...523A..48J}, valid in the range
$-0.4 < \VI < 6$:
\begin{multline}
\BPRP= - 0.0660 + 1.2061\,(\VI) \\- 0.0614\,(\VI)^2 + 0.0041\,(\VI)^3 ~,
\end{multline}
\begin{multline}
G-V= - 0.0257 - 0.0924\,(\VI) \\- 0.1623\,(\VI)^2 + 0.0090\,(\VI)^3 ~.
\end{multline}

\item Determine the spectral type using the effective temperature
approximation \citep{2019hsax.conf..548C}:
\begin{multline}
T_{\rm eff} = 5040\,\Big[ 0.43547 + 0.55278\,(\BPRP) \\
 - 0.046397\,(\BPRP)^2 \Big]^{-1}\quad\units{K} ~.
\end{multline}

\item Compute the apparent \Gband and \Vband magnitudes of the stream stars, neglecting the dust extinction, and correcting the
generated \Gband magnitude and the previously computed \Vband magnitude
from the heliocentric distance of the globular cluster $r_{\rm gc}$, to the
heliocentric distance of the stream star.
\end{enumerate}

\begin{figure}
\includegraphics[width=1.0\columnwidth]{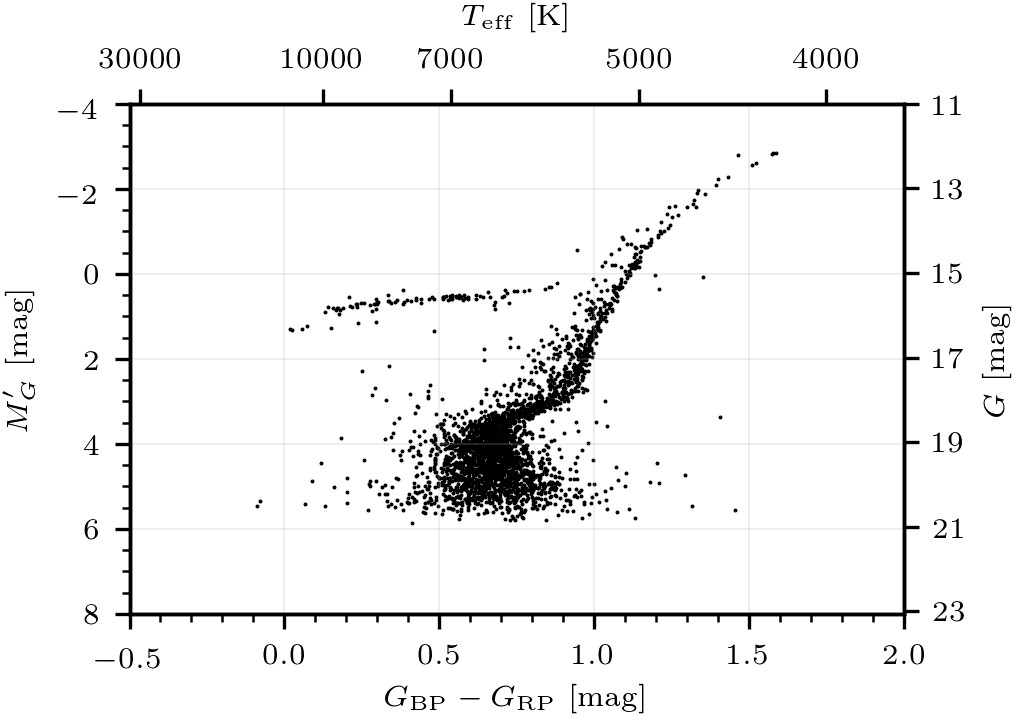}
\caption{Observed $\BPRP$ colour index and \Gband
absolute magnitude $M_G'$ for all stars measured in M68 in GDR2.}
\label{HR_M68}
\end{figure}

\noindent The PyGaia functions give the end-of-mission \textit{Gaia} errors, so
we correct them by multiplying by the scale factors $\lambda_\mu$
(equation\ \ref{eqlam}). The simulated observed coordinates of stream
stars, $w^\mu_o$, are generated by adding a random Gaussian variable with
dispersion $\lambda_\mu\varepsilon_\mu$ to the true coordinates. The
GDR2 measurements generally cannot be performed on stars fainter than
$G=21$, so only stars $G \leqslant 21\, {\rm mag}$ are included in our
simulated catalogue of stream stars. In addition, the small number of
simulated bright stars satisfying
$4 \leqslant G \leqslant 13 \,{\rm mag}$ and
$3550 \leqslant T_{\rm eff} \leqslant 6900 \,{\rm K}$ are given a
radial velocity with the \textit{Gaia} observational error.

 The resulting quantity of simulated stream stars after these cuts is
\mbox{33 228}, and is shown in the top row of Table \ref{res_pre} as the total
number of stream stars in our simulated catalogue before we apply a
number of pre-selection cuts that we describe next.

\subsection{Data pre-selection}\label{data_pre}

 We now imagine that we have a catalogue with all the stars in the
GOG18 simulation (a total of $\Sim 1.5$ billion), and that includes in
addition some fraction of our \mbox{33 228} simulated stream stars. Our method
needs to detect the presence of the stellar stream in an optimal way and
to identify the most likely candidate members.

 To speed up the computation of the likelihood function and reduce the
number of foreground stars of the final selection, most of which are in
sky regions of high stellar density and far from the possible stream
associated with M68, we apply a set of pre-selection cuts to the star
catalogue. The stars that fulfil the following conditions are
pre-selected:
\begin{enumerate}
\itemsep1.0em
\item[(1)] $G \leqslant 21 \,{\rm mag}$, to eliminate faint stars with large astrometric
 errors.
\item[(2)] $\pi < 1/0.3\, {\rm mas}$, e.g. large distance, to eliminate foreground disc stars.

\item[(3)] $|b| > 15\, {\rm deg}$, to avoid the regions of high stellar density close to the disc 
\end{enumerate}

  For our fourth condition (4), we use a more complex pre-selection method to further
restrict the number of stars that are used to evaluate the likelihood
function for different stream models.
The goal is to eliminate most of the foreground stars
that can be ruled out as members of any possible stream associated with
M68, within the uncertainties of the M68 orbit arising from observational
errors and the Galactic potential. Our pre-selection method is described
in detail in Appendix \ref{App6_met}; here we explain its basic idea,
which uses the fact that the tidal stream is close to the orbit of the
progenitor. We use a range of parameters for the orbit of
M68 and the potential of the Milky Way to compute a bundle of possible
orbits of M68 during the time interval from $t_0-t_l$ to $t_0+t_l$,
where $t_0$ is the present time and $t_l$ is chosen to obtain the
relevant part of the orbit for the stream. Our simulations show that
$t_l=50\, {\rm Myr}$ results in an adequate coverage of the reliable
part of the stream in the case of M68, and we adopt this value in this
paper. Many of the stars that are released by the cluster in the first
few orbits, when the rate of escaping stars has not yet settled to a
steady state, are located further from the globular cluster than this
section of the orbit, and they are eliminated from the final catalogue
in this fourth pre-selection condition.

 The bundle of orbits computed in this way is used to define a region in
phase space where stars have to be located to be pre-selected, also taking
into account the observational errors. This is done by
characterizing the bundle of orbits by a series of Gaussians, which are
convolved with the Gaussians of observational errors. Stars have to be
located within this region (involving conditions of the sky positions,
proper motions, and parallaxes) to be pre-selected.

As the final fifth step (5), we remove stars that are within
an angular distance of the globular cluster that gives rise to the tidal
tail, and also stars in any other globular clusters that are within the
pre-selected region, since they have highly correlated kinematics. The
list of globular clusters that have been removed in this way are shown
in Appendix \ref{App6_m68}.

 The number of stars in GOG18 and in the simulated stream of M68 after
each one of these cuts is specified in Table \ref{res_pre}. The first
three steps reduce the general GOG18 catalogue by a factor close to
six, and a smaller factor for the stream stars which are all
sufficiently far from us and mostly at high Galactic latitude. The
fourth step achieves the largest reduction in the number of foreground
stars, a factor $\Sim 400$, by restricting the stars we look for to be consistent with our range of models in sky position, proper motion, and parallax.
The stream stars are reduced by a factor of nearly 3, mostly due to
the stars ejected in the first few orbits in the simulation that are
near the edge of the stream, with a distribution that we consider as
insufficiently reliable. The fifth step eliminates a very small
fraction of stars, and is important mostly to remove stars that are
bound or very close to M68.

\begin{figure*}
\includegraphics[]{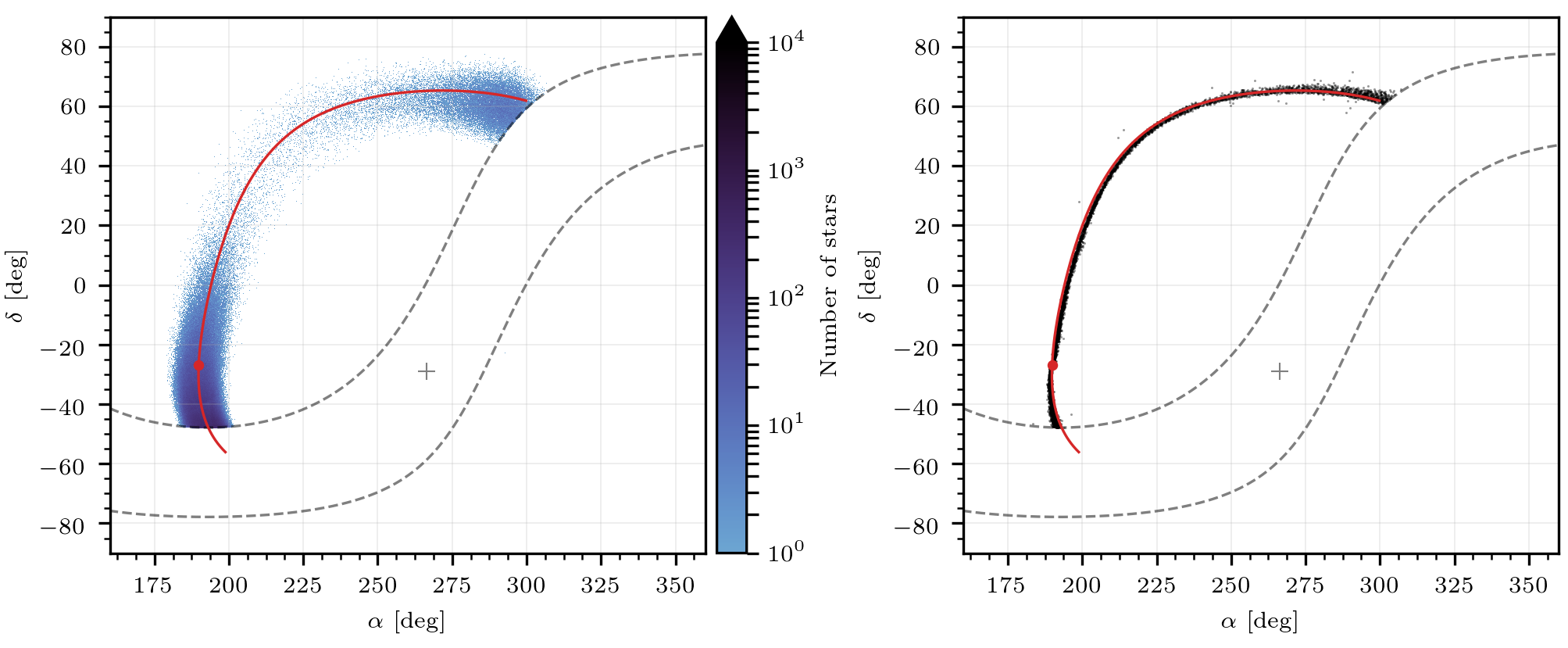}
\includegraphics[]{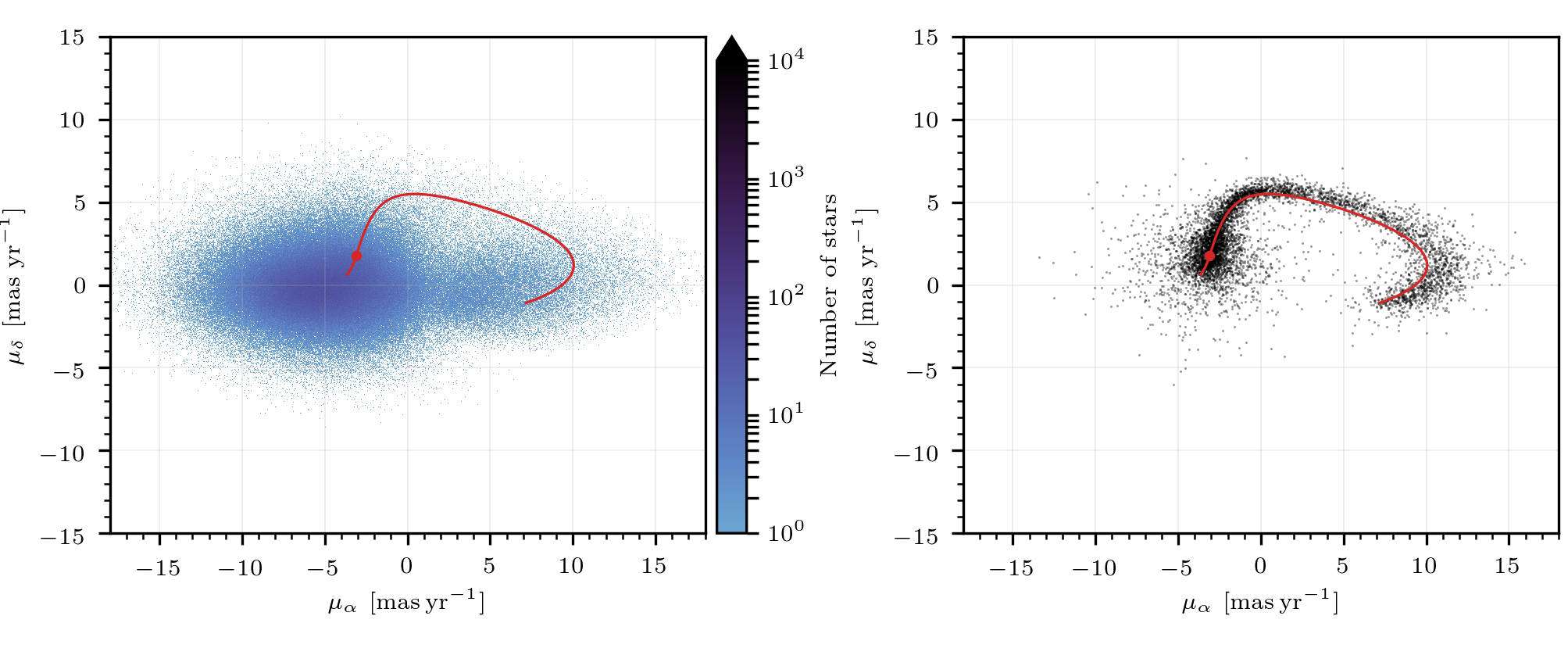}
\caption{\textit{Left}: Distribution of the pre-selected stars from the
GOG18 catalogue. \textit{Right}: Pre-selected stars from the M68
simulated tidal stream. The red dot marks the current position of the
cluster and the red line is its simulated orbit over $t_l=50$ Myr
backwards and forwards from its current position. Grey dashed
lines mark the Milky Way disc cut at $b=\pm15$ deg and the grey cross
indicates the Galactic centre.}
\label{gog18_pre_short}
\end{figure*}

  The distribution of pre-selected stars in the GOG18 catalogue is shown
as blue dots in the left-hand panels of Fig. \ref{gog18_pre_short}, and
the pre-selected stream stars are shown as black dots in the right-hand
panels. Top panels show angular positions and bottom panels show
proper motions. The tidal stream of M68 is particularly favourable to
be observed because of the long stretch of the orbit that is close to
us in a region of low foreground stellar density in the position and
proper motion space.

\begin{table}
\caption[]{\small{Number of stars in the simulated catalogue and simulated
stream that pass our successive pre-selection cuts.}}
\begin{center}
\begin{tabular}{llrr}
\toprule
\multicolumn{2}{l}{\textbf{Pre-selection cut}}&\textbf{GOG18}&\textbf{Stream stars}\\
\midrule
\multicolumn{2}{l}{All catalogue}&1510 398 719&33 228\\
\multicolumn{2}{l}{(1)\,-\,(2)\,-\,(3)}&269 125 739&17 183\\
(4) &&613 098&6627\\
(5) &&612 909&6564\\
\bottomrule
\end{tabular}
\end{center}

\label{res_pre}
\end{table}

\subsection{Recovery of the dark halo parameters}\label{sec:Rec}

  We now test the detection of a tidal stream in the GOG18 simulated
data set by adding a number of stars $N_{\rm str}$ selected randomly among
the 6564 stars of our simulated tidal stream. We find the maximum of
the $\Lambda$ function described in Section \ref{likelihood}
(equation \ref{Lamf}), with the prior in equation (\ref{prif}) that incorporates
the measurements of the M68 kinematics, when the parameters $\tau$
(fraction of tidal stream stars in the data set),
$\theta_c$ (orbital parameters of M68), and $\theta_\phi$ (gravitational
potential parameters) are allowed to vary. Each time we evaluate the
likelihood for a new set of parameters, we need to resimulate the tidal
stream and calculate the probability density of the stars with equation
(\ref{pstream}). To make this computationally easier, we calculate this
stream probability density only for the fixed set of pre-selected stars
described earlier, and we recompute orbits for the tidal stream only for
a fixed set of 1200 test particles in the M68 globular cluster. These
particles are selected among the ones with initial velocities
$v>v_{\rm lim}$ and $r>r_{\rm t}$ (see equations \ref{vlim} and \ref{eqrt})
using a mean cluster orbital radius $R_{\rm c}=21$ kpc in
equation (\ref{eqrt}), which increases the fraction of escaped stars and
the efficiency of the calculation. Typically, the number of stars among
these 1200 that escape M68 and form the tidal stream is about 1000, and
is always greater than 750. We use these stream stars to recompute the
smoothed phase-space density model of the tidal stream with equations
(\ref{pstream}) and (\ref{integral_eq}), for each pre-selected star.

 We find these number of stream stars is sufficient to obtain a
reasonable accuracy for the best-fitting tidal stream. 
We note that choosing fixed initial conditions within
the M68 Plummer model for the simulated stream stars as we vary the
model parameters is important to ensure differentiability of the
final stream star positions and velocities and a smooth likelihood
function when we vary model parameters.

  We compute the maximum likelihood and find the best-fitting stream model
for a total of 30 cases: for six values of the number of stream stars
added to the catalogue, $N_{\rm str}=\{10, 50, 100, 300, 500, 1000\}$, we do
five independent cases with different random selections of $N_{\rm str}$
stars among all the 6564 selected escaped stars in our base
simulation of the M68 tidal stream (bottom row of Table \ref{res_pre}).

\begin{figure}
\includegraphics[width=1.0\columnwidth]{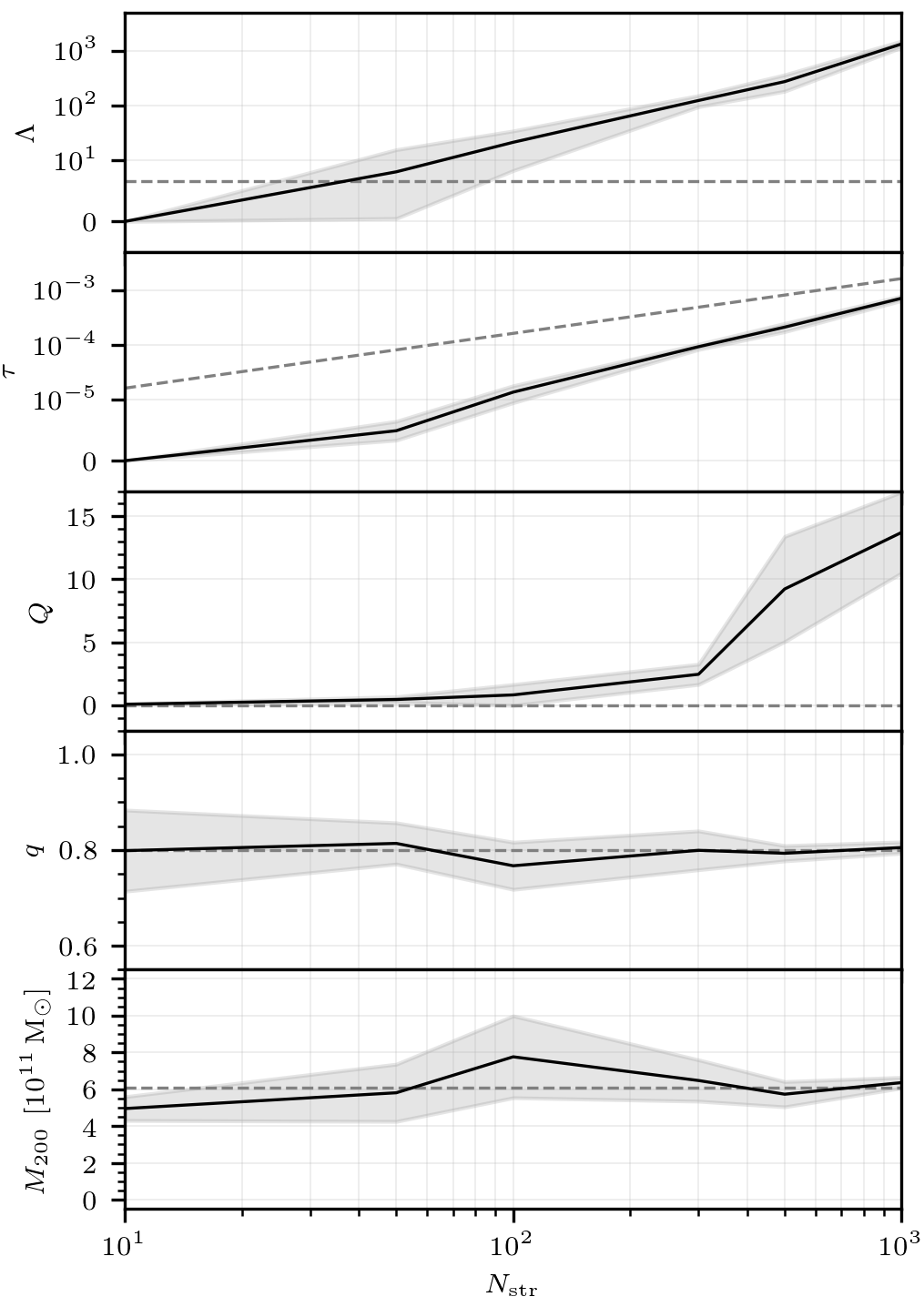}
\caption{Computed statistical parameters and best estimates of the halo
mass and axial ratio using the GOG18 pre-selected data and five different
random samples of sets of $N_{\rm str}$ simulated stream stars, with
$N_{\rm str} = \{10, 50, 100, 300, 500, 1000\}$. The first grey dashed line
marks the threshold $k=6.635$, and the next lines the true parameter
values that were used in the tidal stream simulation
(see Table \ref{M68_DH}). Solid lines show the average best fit obtained
for each $N_{\rm str}$, with the range shown by the grey band.}
\label{res}
\end{figure}

  Results are shown in Fig. \ref{res}. In each panel, the solid line
connects the average results for the five cases of each value of
$N_{\rm str}$, while the grey band is their range. The top panel shows the
value of $\Lambda$. When $\Lambda > k=6.635$ (shown as the horizontal
dashed line), the detection of the stream is significant at the
$\epsilon=0.01$ probability of rejecting the null-hypothesis. This
happens always when $N_{\rm str}>100$, and in most cases for $N_{\rm str}>50$.

  The second panel is the fraction $\tau$ of stream stars in the
catalogue. The recovered fraction is generally lower than the true value
(equal to the ratio of added stream stars to the total number of
pre-selected stars), indicated by the dashed line.
In general, $\tau$ can be different than the true value for two
reasons: our foreground model is highly approximate and does not
accurately reflect the distribution of stars in GOG18, and the stream
phase-space density model we construct by adding Gaussians is also
imprecise. The latter may explain the increasing ratio of the measured
to the true $\tau$ with $N_{\rm str}$, if the algorithm tends to match the
positions of a fraction of the stream stars while ignoring the rest at
low $N_{\rm str}$.

  The third panel is the value of $Q$ from equation (\ref{eqqu}). This
value is small when the detected stream does not impose
significantly improved constraints on the globular cluster orbit
compared to the prior (the measured proper motions, radial velocity, and
distance), but grows as the stream stars provide more stringent
constraints on the orbit. If $Q>4$, the detected stream is pushing the
best-fitting orbit away from the measured values. Significant deviations
start to occur in our simulations for our largest value of $N_{\rm str}$,
probably due to the approximate evaluation of the tidal tail phase-space
distribution in our method. These deviations are not so large as to
substantially affect our best fits.

  Finally, the last two panels show the recovered values of the dark
halo mass and its axial ratio. These recovered values are consistent
with the true ones, and the errors are as small as $\Sim 10$ per cent for
the mass and $\Sim 3$ per cent for the axial ratio when $N_{\rm str} \Sim 1000$.
Of course, realistic errors on these potential parameters are expected
to increase when allowing for variations in other parameters such as
the disc mass and scale length or the halo density profile.

\subsection{Identification of stream stars}\label{sec:ids}

 Our final goal is to identify the stars that are most likely to be
members of the stream among our simulated stars, and check if the true
stream stars that were inserted in the catalogue are recovered. We first
identify stars that are consistent with a phase-space model of the tidal
stream and then we select those that are compatible with the colour and
magnitude of the globular cluster progenitor.

\subsubsection{Phase-space identification}

 We start by considering only the phase-space variables. We construct
again the phase-space density model of the stream with the same
procedure as in Section \ref{simuTS}, using the estimated best-fitting
parameters $\hat{\theta}_\kappa$ and increasing now the number of
simulated escaped stars to $N_{\rm e}=10^4$, which yields a more accurate and
smoother model than the smaller number used when the model parameters
are varied. We compute the probability density of a star to belong to
the tidal stream with equation (\ref{integral_eq}), although without
considering this time the selection function $\psi_{\rm s}$,
\begin{equation}
P_\SM{SEL}\var{w^\mu|
\hat{\theta}_s,\hat{\theta}_c,\hat{\theta}_\phi;\sigma^{\mu\nu}} =
 \frac{1}{N_{\rm e}} \, \sum_{i=1}^{N_{\rm e}}
 G\var{w^\mu-w^\mu_{{\rm c}i}|\sigma^{\mu\nu}\!+\Xi_i^{\mu\nu}} ~.
\end{equation}
We identify as candidate stream members the set of stars that pass a sixth
cut (6), requiring a probability density above a fixed threshold:
\begin{equation}
 P_\SM{SEL} \geqslant \chi_{\rm sel} ~.
\end{equation}
The threshold $\chi_{\rm sel}$, with units of the inverse product of the
six $w^\mu$ coordinates that we shall express in ${\rm yr}^{3} \, {\rm deg}^{-2} \, {\rm pc}^{-1} \, {\rm mas}^{-3}$, can be
conveniently chosen to retain as many candidates as possible without
excessively contaminating the sample that is obtained with foreground
stars.

\subsubsection{Colour-magnitude selection}\label{col_sel}

 As the final condition to consider a star as a candidate member of a
stellar stream, we consider the colour information that is obtained in
the \textit{Gaia} photometric measurements. A stellar stream member should have
colours and absolute magnitude (which can be computed assuming the model
distance of the stellar stream) consistent with the HR-diagram of the
progenitor cluster.

 The GOG18 catalogue includes a simulation of the \textit{Gaia} photometric
measurements of the \Gband magnitude, roughly corresponding to
unfiltered light over the wavelength range from $\Sim 330$ to 1050 nm.
Two additional magnitudes are also measured in a blue (BP) and red (RP)
broad passbands from $330$ to $680$ nm, and from $630$ to $1050$ nm,
respectively, yielding the two magnitudes $G_{\rm BP}$ and $G_{\rm RP}$. To
compute the absolute magnitude $M_{G}'$, we do not use \textit{Gaia} parallaxes
because the observational errors are too large. Instead, we assign to
each star the heliocentric distance of the closest point to the star of
the computed orbit of the progenitor cluster, in our model of the
Galactic potential that has given the best-fitting stellar stream. We then
select the stars with colours and absolute magnitude that are consistent
with the HR-diagram of the progenitor cluster, following the procedure
that is described in detail in Appendix \ref{App7}.

 In the case of M68, the tidal stream is expected to pass at $\Sim 5$
kpc from the Sun, so the detectable stream stars close to us should
often have lower luminosity than the least luminous detectable stars
at the M68 distance of $\Sim 10$ kpc. These stars cannot be directly
compared to the M68 HR-diagram as measured by \textit{Gaia}. We solve this
problem by including also as candidate stars those with absolute
magnitude $M'_G \geqslant 5.68$, and colour $0.5 \leqslant \BPRP
\leqslant 1$ mag, which adequately brackets the main-sequence for stars
in the relatively narrow range of distances the M68 stream extends
over. We neglect any impact of dust extinction, which is small and
fairly constant along the tidal stream according to the
STILISM\footnote{\url{https://stilism.obspm.fr/}} extinction model
\citep{2017A&A...606A..65C, 2018A&A...616A.132L}. In summary, our
seventh selection cut (7) is that the star colours and absolute magnitude
are either compatible with the M68 HR-diagram observed by \textit{Gaia}, or obey
the above conditions for main-sequence stars in M68 of lower luminosity
than the \textit{Gaia} detection threshold.

\begin{figure}
\includegraphics[width=1.0\columnwidth]{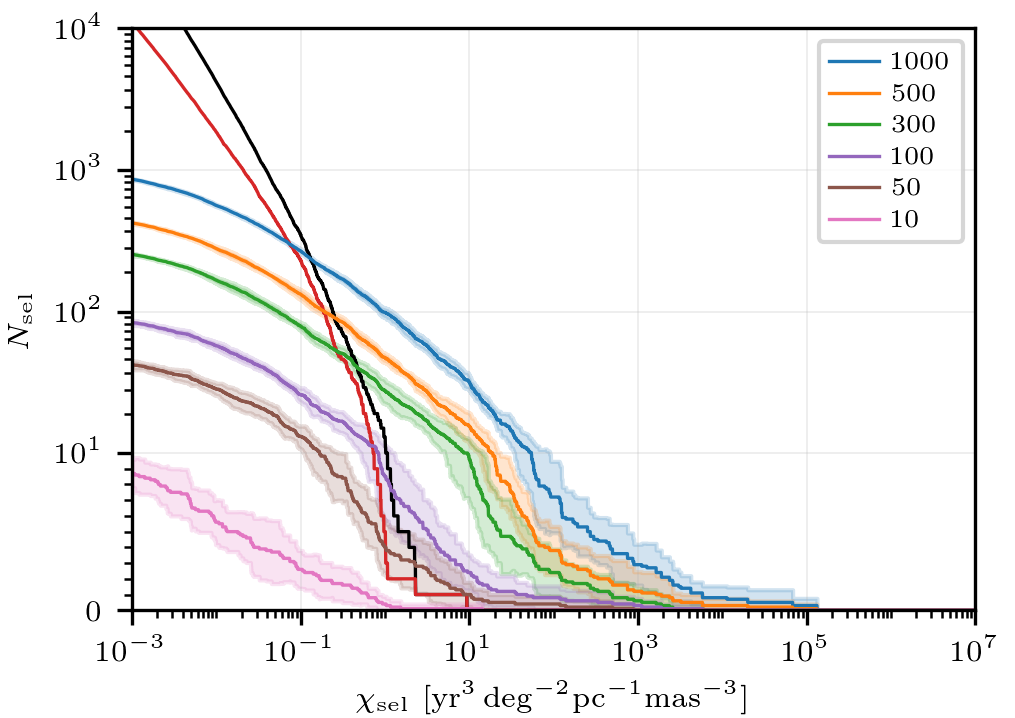}
\caption{Number of selected stars $N_{\rm sel}$ as a function of the
selection threshold $\chi_{\rm sel}$ in our sixth cut. The black line
indicates the number of GOG18 stars compatible with the phase-space
model of the stellar stream (cuts from 1 to 6 applied), and the red line
shows stars that are additionally colour-magnitude compatible with M68
(passing the seventh cut as well). Colour lines are the number of stream
passing all seven cuts from six different cases with $N_{\rm str}$ stars in
our model stream added, with their range in several random samples
indicated by the shaded areas.}
\label{sel_simu}
\end{figure}

 The results of our simulations where we add a randomly selected set
of $N_{\rm str}$ stars of the simulated stream to the GOG18 catalogue
is presented in Fig. \ref{sel_simu}. The number of selected stars
$N_{\rm sel}$ from GOG18 alone after our first six cuts are applied is
the black line, shown as a function of the selection threshold,
$\chi_{\rm sel}$. The red line is the number of stars that are in addition
colour-magnitude compatible with M68 (cut 7). The other coloured lines
show the number of selected stars when we add $N_{\rm str}$ simulated
stream stars to GOG18, taking several random samples of these added
stars to compute a mean of $N_{\rm sel}$ and its dispersion (shown as
the shaded area around each curve). For each case, we have constructed the phase-space
model of the tidal stream using the computed values of the parameters
shown in Fig. \ref{res}. The added stream stars pass all of our seven cuts,
although the seventh cut in this case is automatically satisfied because the
model stream stars are assumed to have compatible colours with M68. The
figure shows that for the M68 simulated stream, and by choosing a
threshold $\chi_{\rm sel}\sim$ 3 ${\rm yr}^3 \, {\rm deg}^{-2} \, {\rm pc}^{-1} \, {\rm mas}^{-3}$ , we select $\Sim 10$ per cent of the stream
stars measured in the catalogue while including only 1 foreground star
among the selected ones. This performance improves if we restrict our
selection to phase-space regions of low background contamination, and
is also sensitive to the way the selection function $\psi_{\rm s}$ is treated
(which has been ignored here).

 Fig. \ref{sel_simu} also shows that the use of the colour information
is not essential to the ability to find the stellar stream, although it
certainly helps to constrain further the member stars as quantified by
the separation between the black and red lines. As long as the number of
stars in the stream is $N_{\rm str} \GtrSim 100$, the stream is detected
as an excess of stars above the probability threshold $\chi_{\rm sel}$,
and when colours are used this minimum number of required stream stars is
reduced to $N_{\rm str}\simeq 50$, assuming that the orbit of the
progenitor cluster is known. This, of course, varies for each candidate
cluster progenitor, depending on the level of foreground contamination
of the zones covered by the predicted stellar streams and the complexity
and breadth of the predicted stellar stream.

\section{Application to M68 using \textit{G\lowercase{aia}} DR2 data}\label{sec4}

\subsection{Data pre-selection}

 The full sky GDR2 star catalogue, published on 2018 April 25 based on data collected during the first 2 yr
of the \textit{Gaia} Mission \citep{2016A&A...595A...1G}, includes five-parameter
astrometric solutions (parallaxes, sky coordinates, and proper motions)
and multiband photometry ($G$, $G_{\rm BP}$, and $G_{\rm RP}$ magnitudes) of
$\Sim 1.7$ billion sources. In addition, it includes radial velocities
for $\Sim 7.2$ million sources. A complete description of its contents
is found in \citet{2018A&A...616A...1G}.

 We apply the pre-selection method described in Section \ref{data_pre}
to GDR2, as well as to the GOG18 simulated catalogue to compare results.
The numbers of stars that pass each of our cuts are given in Table
\ref{results_cuts}. The first three cuts produce a number of selected
stars similar in GOG18 and GDR2. The fourth cut, requiring stars to be 
close to the predicted M68 orbit for a variety of models, results in the 
largest reduction. The number of stars in GDR2 after this cut is smaller 
than in GOG18, which we think is attributable to imperfections in the 
model of the stellar distribution of the GOG18 simulation and our 
approximate treatment of the effect of measurement errors in GDR2. The 
fifth step results in a larger reduction of GDR2 compared to GOG18, 
because of the presence of stars belonging to the globular cluster M68
(which are not actually simulated in GOG18), although the reduction is 
still small.

\begin{figure*}
\includegraphics[scale=0.96]{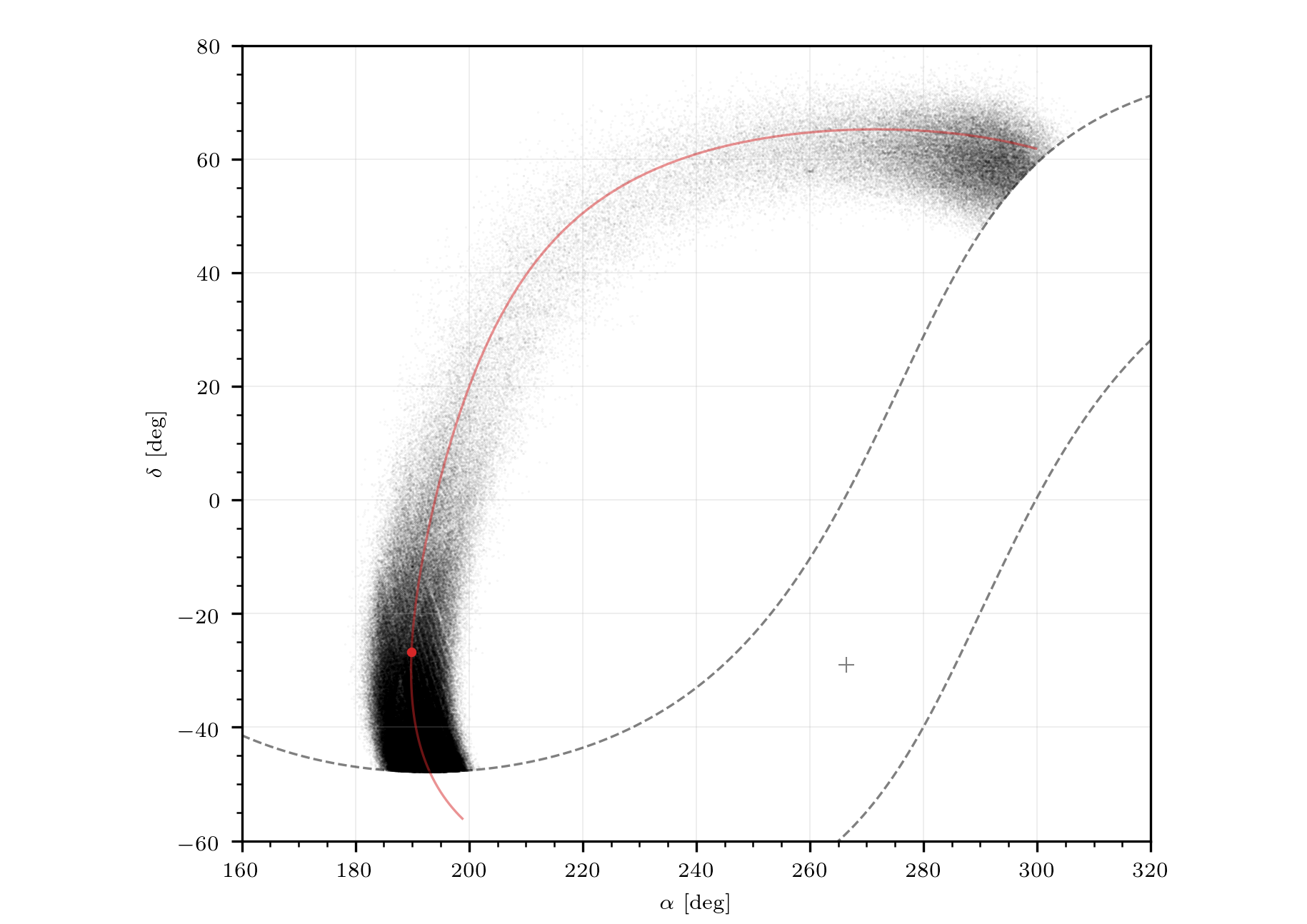}
\caption{Sky map in equatorial coordinates of pre-selected stars passing
our first five cuts. Grey dashed lines indicate a Galactic latitude
$b=\pm15$ deg and the grey cross is the Galactic centre. Red dot marks
the present position of the cluster M68 and the red curve is its predicted orbit over 50 Myr forwards
and backwards respect its current position.}
\label{ad0_sel}
\end{figure*}

 As a first exploration of the GDR2 pre-selection, we plot the sky
coordinates of the pre-selected stars. This is shown in Fig.
\ref{ad0_sel} for all the \mbox{440 499} stars after our cut 5, plotted as very
small black dots. The large red dot indicates the position of M68 and
the light red curve is its computed orbit in our standard model (central
parameters in Table \ref{M68_DH}). The cross shows the Galactic centre
and the two dashed lines indicate a Galactic latitude
$b=\pm 15\, {\rm deg}$. Without our more strict selection from cut 6,
the presence of any tidal stream is not clearly discerned owing to the
large foreground. Fig. \ref{ad0_sel} shows,
however, the presence of variations of stellar density in the form of
parallel streaks due to the \textit{Gaia} exposure variations, most clearly
visible in the range $-30 \,\, {\rm deg} < \delta < 0 \,\, {\rm deg}$. There are also
regions of low density that run parallel and close to the predicted
M68 orbit, which is a reason to be concerned with our method of
identifying a tidal stream.

\begin{figure*}
\includegraphics[scale=0.96]{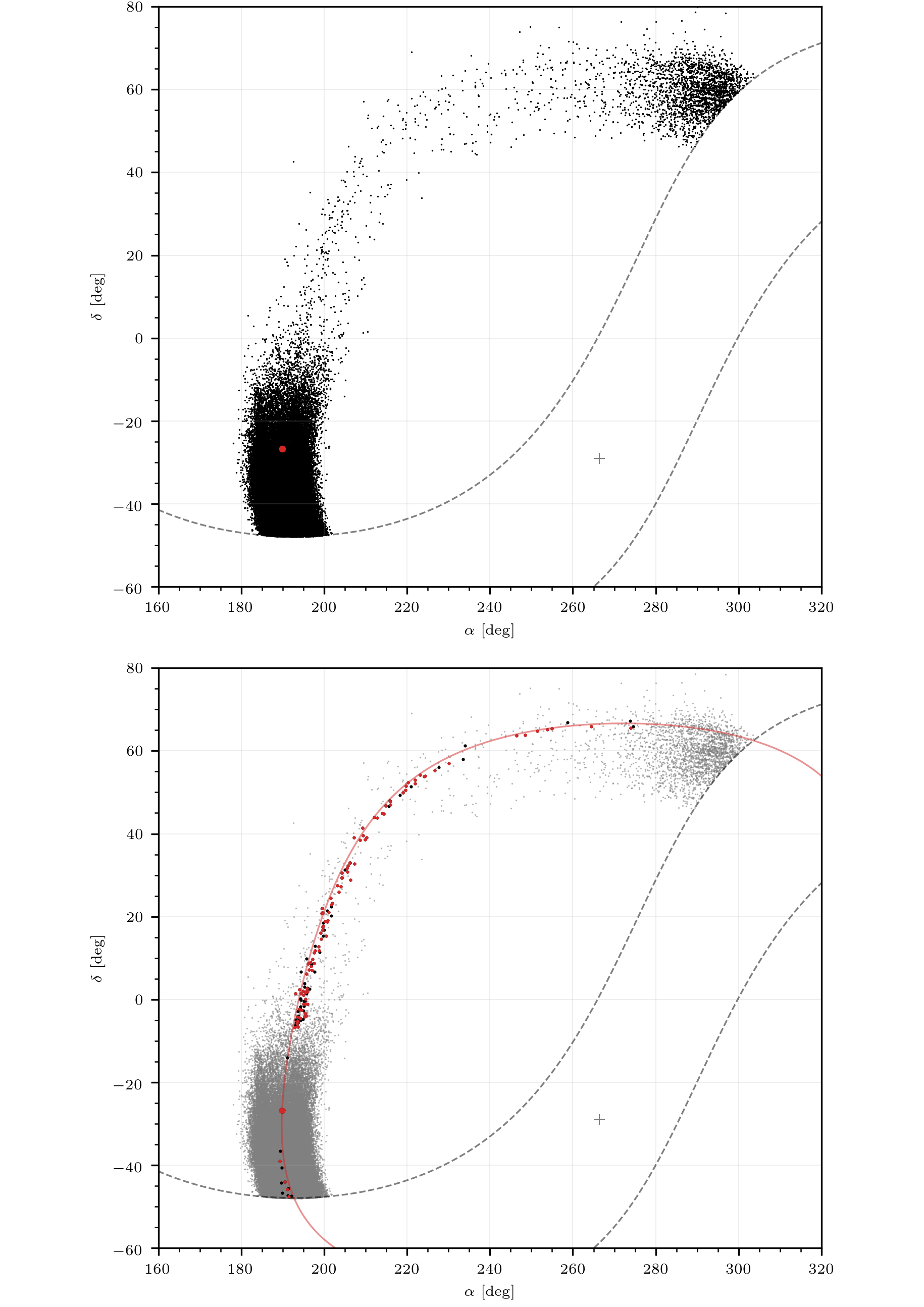}
\caption{
\textit{Top}: Same as Fig. \ref{ad0_sel}, but restricted now to stars
that pass our seventh cut in addition to the first 5 (i.e., with photometry
compatible with the M68 HR-diagram), which leaves $\Sim 30$ per cent of the
stars in Fig. \ref{ad0_sel}.
\textit{Bottom}: Final selection of stars. Grey small dots are the same
as in top panel, black dots are stars compatible with the density model
of the stellar stream (cut 6) and red dots are stars that are also
compatible with the M68 HR-diagram (cut 7). The small number of selected
stars in the range $-50\,\, {\rm deg} \leqslant\delta\leqslant-7 \,\, {\rm deg}$ and the abrupt
change for $\delta > -7$ deg is caused by our selection method.}
\label{ad_sel}
\end{figure*}

 To see if the stream can be more easily identified using only our
broad pre-selection in cut 4 when we include the colour information,
we apply now the extra cut 7 defined in Section \ref{col_sel} and described
in detail in Appendix \ref{App7}, to select stars with \textit{Gaia} colours
compatible with the M68 HR-diagram. This reduces the number of
pre-selected stars to \mbox{127 615}. The positions of these stars are plotted
as black dots in the top panel of Fig. \ref{ad_sel}. The expected
elongated overdensity of stars along the predicted orbit of the globular
cluster is now more clearly discerned extending over a large part of the
North Galactic hemisphere.

\begin{table}
\caption[]{\small{Total number of stars in GOG18 and GDR2 and
number of stars that pass each cut. For the last two cuts 6 and 7, this
is shown for three separate sky regions defined in Section \ref{sel_ss}, in
which the tidal tail is detected in different foreground conditions.
Numbers in parentheses for GOG18 indicate the expected number of stars
in the absence of any tidal tail if GOG18 had the same number of
pre-selected stars as GDR2 in each of the three regions.}}
\begin{center}
\begin{tabular}{llrr}
\toprule
\multicolumn{2}{l}{\textbf{Pre-selection cut}}&\textbf{GOG18}&\textbf{GDR2}\\
\midrule
\multicolumn{2}{l}{All catalogue}&1510 398 719&1692 919 135\\
\multicolumn{2}{l}{(1)\,-\,(2)\,-\,(3)}&269 125 739&276 019 797\\
(4) &&613 098&446 982\\
(5) &&612 909&440 499\\
\midrule

\multicolumn{2}{l}{\textbf{Region (i) Circle}}&&\\
\midrule
(6) $\chi_{\rm sel} = 0.554$&&1 (1)&13\\
(7) &&1 (1)&13\\
\midrule

\multicolumn{2}{l}{\textbf{Region (ii) Disc 1}}&&\\
\midrule
(6) $\chi_{\rm sel} = 1.392$&&6 (4)&12\\
(7) &&2 (1)&4\\
\midrule

\multicolumn{2}{l}{\textbf{Region (iii) Halo}}&&\\
\midrule
(6) $\chi_{\rm sel} = 5.65\pd{-3}$&&17 (33)&126\\
(7) &&1\phantom{7} (2)&98\\
\midrule
\multicolumn{2}{l}{\textbf{Final selection}}&&\\
\midrule
(6) &&24 (38)&151\\
(7) &&4\phantom{4} (4)&115\\
\bottomrule
\end{tabular}
\end{center}

\label{results_cuts}
\end{table}

\subsection{Dark halo parameters}

 The values of the model parameters maximizing the likelihood ratio
$\Lambda$ (which includes the likelihood function of the stream and the
kinematic measurements of M68, eqs.\ \ref{Lamf} and \ref{prif}) have
been calculated applying the Nelder-Mead Simplex algorithm to the GDR2
pre-selected data after our first five cuts. A total of nine parameters are
varied: the fraction of pre-selected stars $\tau$ in the stellar stream,
the four parameters $\rho_{0 \rm dh}, a_{1 \rm dh}, a_{3 \rm dh}$, and $\beta_{\rm dh}$ of the halo
gravitational potential and the distance and three velocity components
of M68. In this case, we have computed only the diagonal elements of the
covariance matrix in equation (\ref{uneq}) and used them to compute
errors of the parameters assuming that all the other ones remain fixed.
The results are listed in Table \ref{results_param}.

The main conclusions we infer from these results are the following:
\begin{itemize}
\item The maximum value found for the likelihood ratio statistic is
$\varLambda = 84.6$, implying that the null-hypothesis ($\tau = 0$) is
rejected at high confidence because $\varLambda > k = 6.635$ represents
the 99 per cent confidence level.
\item Comparing with the simulation results in Fig. \ref{res}, we find
that the value $\varLambda \sim 80$ suggests that there are
$N_{\rm str} \sim 250$ stars that belong to the
detected tidal tail and that the tidal stream would have been detected
as long as the number of stars is $N_{\rm str}\GtrSim 100$ using only the
kinematic data (without using any colour information).
\item The estimated distance and velocity of the globular cluster M68
change little compared to the directly measured values of the proper
motion, distance, and radial velocity (Table \ref{M68_DH}). The reason is
that the detected stream does not constrain the orbit of the cluster
with greater precision than the kinematic measurements of M68
themselves. This shows that the detected stream is fully consistent with
originating in M68.
\item The best-fitting model of the Milky Way dark matter halo has the
parameters shown in Table \ref{results_param}. The implied total
circular velocity at the solar radius is $v_{\rm c} = 225.38$ km s$^{-1}$, compatible
with the circular velocity of the Local Standard of Rest (LSR)
\citep[][see Appendix \ref{App2}]{2016ARA&A..54..529B}. Fig. \ref{rot}
shows the rotation curve of the Milky Way and the contribution of each
component.
\item The value of the halo density minor-to-major axial ratio in our
best-fitting model is $q = 0.87\pm0.06$, corresponding to a potential
flattening along the $z$-axis of $q_{\varPhi} = 0.94 \pm 0.03$. However,
the error has been computed only for our parameterized model and is not
marginalized over the other parameters determining the radial profile.
Constraints on the halo oblateness need to be obtained by considering
possible variations in the mass and radial profile of the disc and the
bulge mass, which is beyond the scope of this paper. We plan to examine
constraints on the shape of the dark matter halo in the future, using
also data for other streams; nevertheless, our model is compatible with
previous studies based on GD-1
\citep{2010ApJ...712..260K,2015MNRAS.449.1391B,2019MNRAS.486.2995M} and
Palomar 5 \citep{2015ApJ...803...80K, 2016ApJ...833...31B}.
\end{itemize}

\begin{table}
\caption[]{\small{Best-fitting results for the orbit of M68 and the dark
halo parameters, using the GDR2 pre-selected data.}}
\begin{center}
\begin{tabular}{lllc}
\toprule
\multicolumn{2}{l}{\textbf{Statistical parameters}}\\
\midrule
$\varLambda$&&$84.6$\\
$\tau$&&$(2.17\pm0.42)\pd{-4}$\\
$Q$&&$0.6723$\\
\midrule
\multicolumn{3}{l}{\textbf{Best-fitting kinematics of M68}}\\
\midrule
$r_{\rm h}$&\units{kpc}&$10.24\pm0.05$\\
$v_r$&\units{km s$^{-1}$}&$-94.544\pm0.052$\\
$\mu_\delta$&\units{mas yr$^{-1}$}&$1.7917\pm0.0020$\\
$\mu_\alpha$&\units{mas yr$^{-1}$}&$-3.0953\pm0.0035$\\
\midrule
\multicolumn{3}{l}{\textbf{Best-fitting parameters of dark halo}}\\
\midrule
$\rho_{0 \rm dh}$&\units{M$_{\odot}$ kpc$^{-3}$}&$(7.268\pm0.076)\pd{6}$\\
$a_{1 \rm dh}$&\units{kpc}&$18.59\pm0.73$\\
$a_{3 \rm dh}$&\units{kpc}&$16.17\pm0.96$\\
$\beta_{\rm dh}$&&$3.102\pm0.039$\\
\midrule
$M_{200}$&\units{M$_{\odot}$}&$(6.37\pm0.35)\pd{11}$\\
$q$&&$0.87\pm0.06$\\
$q_{\varPhi}$&&$0.94\pm0.03$\\
\midrule
\multicolumn{3}{l}{\textbf{Derived orbital parameters of M68}}&\\
\midrule
$r_{\rm peri}$&\units{kpc}&$6.95\pm0.03$&\\
$r_{\rm apo}$&\units{kpc}&$42.3\pm3.4$&\\
$L_{z}$&\units{km s$^{-1}$ kpc}&$-2414.7\pm4.4$&\\
\bottomrule
\end{tabular}
\end{center}
\label{results_param}
\end{table}

\begin{figure}
\includegraphics[width=1.0\columnwidth]{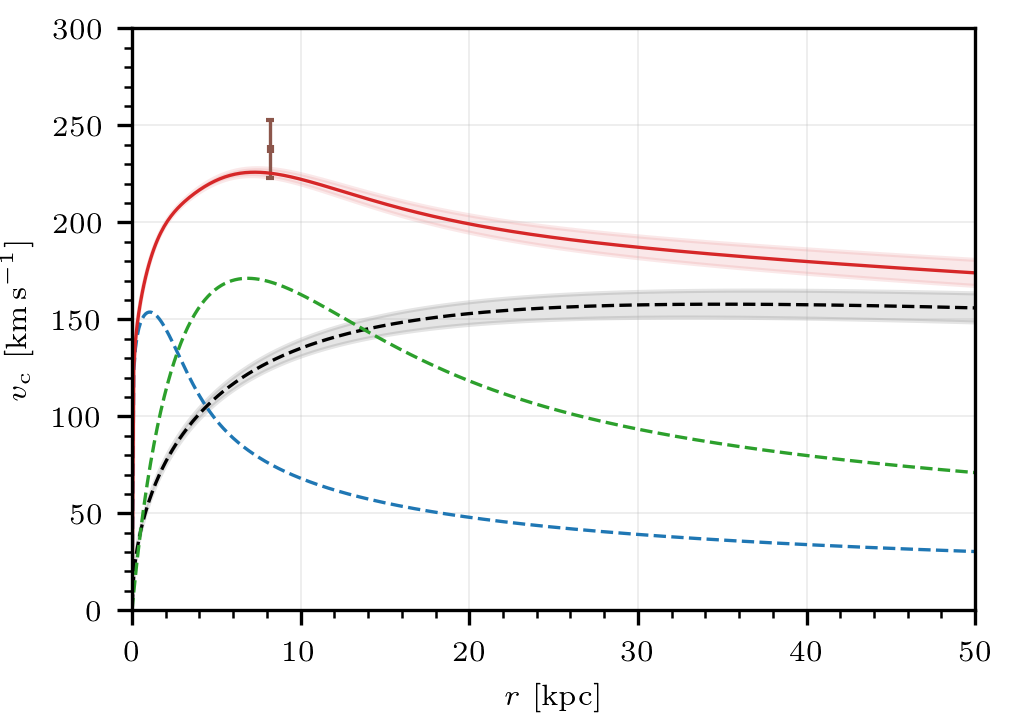}
\caption{Rotation curve of the Milky Way (red). Contributions from
bulge (blue), thin and thick disc (green), and dark halo (black) are
shown as dotted lines. Brown point marks the circular velocity of the
LSR at the Sun's distance.}
\label{rot}
\end{figure}

\subsection{Selection of stream stars}\label{sel_ss}

 We now seek to identify the stars among our pre-selected set of
\mbox{440 499} which have a high probability of belonging to the identified
M68 tidal stream. We do this by applying our sixth cut in a similar way
as with our simulated stream in Section \ref{sec:ids}, choosing the threshold $\chi_{\rm sel}$ that maximizes the ratio between
the GDR2 and the GOG18 selections. This maximum occurs for a low number
of GOG18 selected stars and its value is affected by Poisson
fluctuations of the sample. This selection criterion minimizes the
foreground contamination in the final selected sample, taking only a few
expected foreground stars.

 The foreground stellar density has large variations over the sky area
of the pre-selected sample (after applying our first five cuts), implying
that a single value of $\chi_{\rm sel}$ would not be an efficient way
of obtaining the largest possible sample of reliable stream candidates
while minimizing the foreground contamination. We therefore divide the
pre-selected sample into four sky zones:
\begin{enumerate}
\itemsep1.0em
\item[(i)] \textit{Circle}: A circle of angular radius 0.5 deg
centred on M68.
\item[(ii)] \textit{Disc 1}:
 $\quad -50 \leqslant \delta \leqslant -7 \qquad 180 \leqslant \alpha \leqslant 200 \qquad \units{deg}$
\item[(iii)] \textit{Halo}:
 $\quad\hspace{1.07em} -7 < \delta \leqslant 80\hspace{0.29em} \qquad 180 \leqslant \alpha \leqslant 275 \qquad \units{deg}$
\item[(iv)] \textit{Disc 2}:
 $\quad\hspace{0.53em} 45 \leqslant \delta \leqslant 80\hspace{0.28em} \qquad 275 < \alpha \leqslant 310 \qquad \units{deg}$
\end{enumerate}
The second region called Disc 1 excludes the circle region around
M68. Notice also that the circle region includes stars that are within
0.5 deg, but further than 0.3 deg from the centre of M68 because
of our cut 5 in the pre-selected sample.

\begin{figure}
\vspace{0.20em}
\includegraphics[width=1.0\columnwidth]{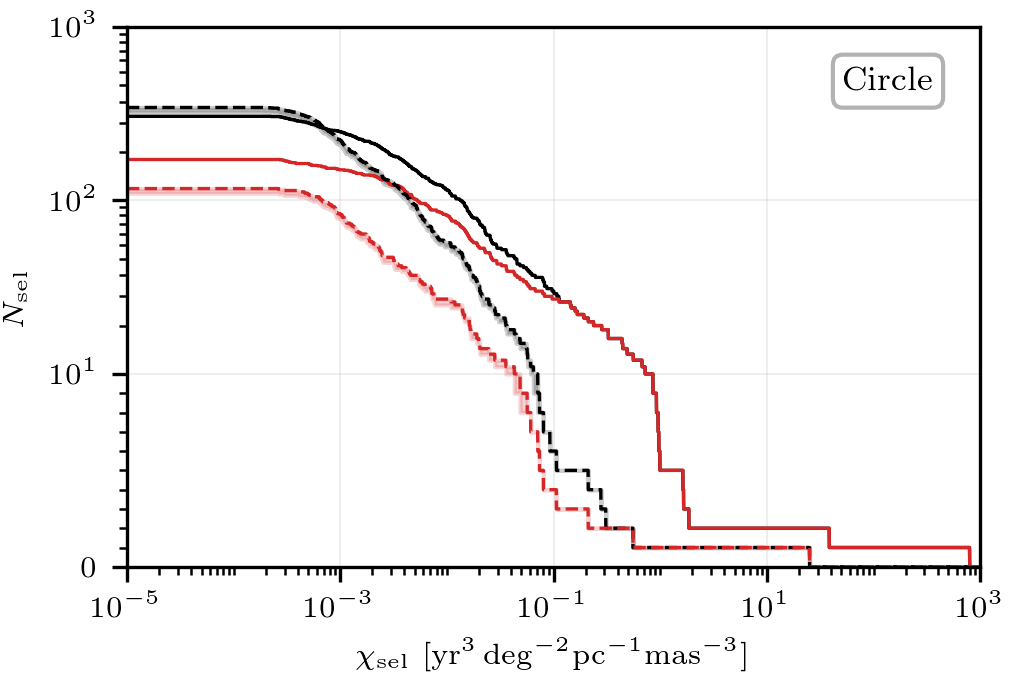}
\includegraphics[width=1.0\columnwidth]{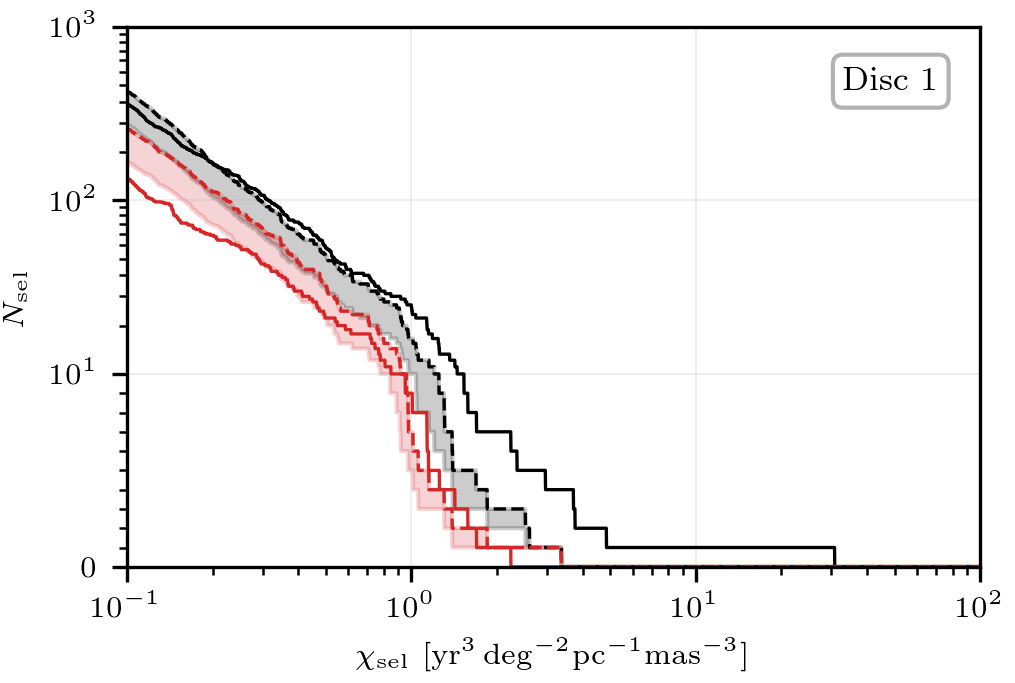}
\includegraphics[width=1.0\columnwidth]{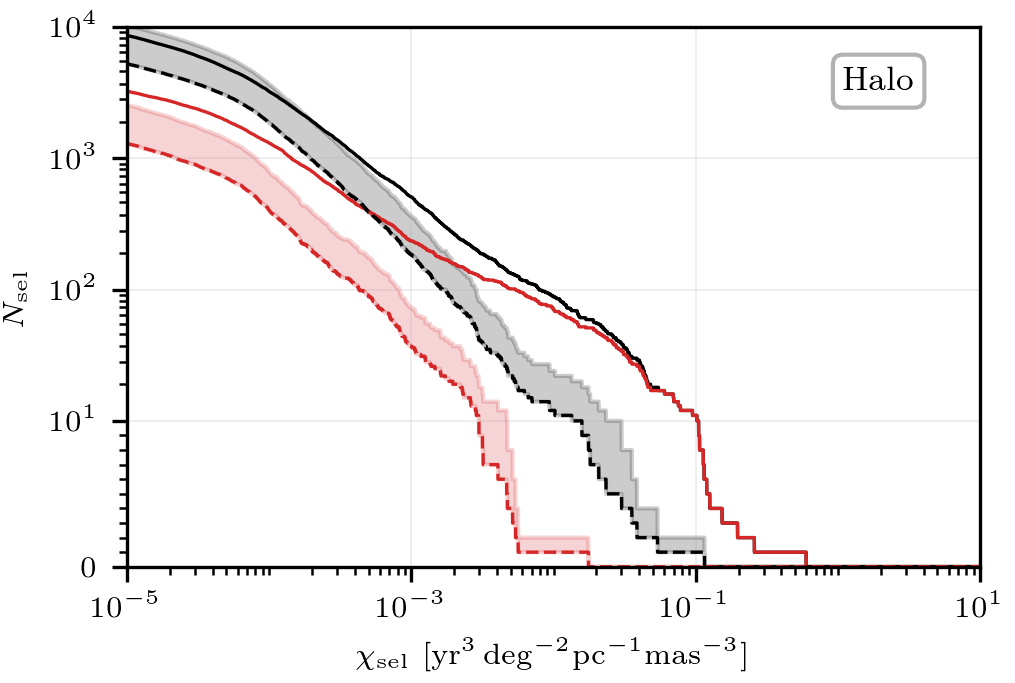}
\caption{Number of selected stars $N_{\rm sel}$ as a function of the
selection threshold $\chi_{sel}$. Black line is the number of GDR2 stars
compatible with the stellar stream phase-space model, red line are stars
additionally compatible with the M68 colour-magnitude diagram. Dashed
lines are the same quantities for GOG18 stars. The shaded space marks
the difference between GOG18 and its correction.
\textit{Top}: Zone (i) Circle. \textit{Middle}: Zone (ii) Disc 1.
\textit{Bottom}: Zone (iii) Halo.}
\label{sel_res}
\end{figure}

 The number of selected stars $N_{\rm sel}$ in the GOG18 and the GDR2
catalogues as a function of the selection threshold $\chi_{\rm sel}$ is
shown in Fig. \ref{sel_res}, as the black dashed and solid lines,
respectively. We show this only for the first three regions (results
for the fourth region, called Disc 2, are similar to the Disc 1 region).
When we apply, in addition, the colour selection cut 7 to require that
the star colours are consistent with the M68 HR-diagram, we obtain the
solid and dashed red lines for GDR2 and GOG18.

  In general, at low $\chi_{\rm sel}$, the number of stars in GOG18 and
GDR2 is not exactly the same: there are more stars in GDR2 compared to
GOG18 in the halo and a similar number (slightly higher in GOG18) in
the other three regions. As discussed previously, we believe this is due
to imperfections of the model used to construct the GOG18 catalogue in
modelling the real Milky Way galaxy and also to approximations we have
used to take into account the effect of astrometric errors in GDR2. At
high $\chi_{\rm sel}$, the number of stars in GDR2 increases compared
to GOG18, mainly in the Circle and Halo regions, as expected if the
tidal stream is real, and from the presence of stars bound to M68. In
the Disc regions, the tidal stream is barely noticed due to the high
foreground contamination. 

 To correct for the different number of foreground stars in the GOG18
simulation and the real GDR2 data, we multiply the number of stars
found in GOG18 to pass the cut 6 in each region by the ratio of the
total number of pre-selected stars in GDR2 to that in GOG18 (passing
the first five cuts). This corrected number is given in parenthesis
in Table \ref{results_cuts} in the GOG18 column. We also give the value
of $\chi_{\rm sel}$ used in each region. In the halo region, we can
afford using a low number because of the very low foreground
contamination, but in the other regions, the threshold needs to be set
to a much higher value to avoid picking up too many foreground stars as
candidates. The number of stars left after applying also the colour cut
7 are also given in Table \ref{results_cuts} for GOG18 and GDR2.

  We find the following results for the stars that are finally selected
as likely members in each zone:

\begin{enumerate}
\itemsep1.0em
\item[(i)] \hspace{0.155em} \textit{Circle}: The selected stars are at
a projected distance of $50$ - $90$ pc from the centre of the
progenitor because of our cut 5 in the pre-selection and the definition
of this region, so they lie in the transition between the cluster and
the stream. The 26 stars with the highest intersection with the stream
density model are colour-magnitude compatible with the cluster and 13
of them are selected above our chosen $\chi_{\rm sel}$. Only one of
these stars should belong to the foreground on average. Many of these
stars, even if unbound from M68, may be orbiting practically at the same
orbital energy and may therefore be moving in loops around M68.
\item[(ii)\phantom{i}] \textit{Disc 1}: This is the most contaminated zone
because it is near the Galactic disc and the proper motions of the disc
stars overlap that of the globular cluster. This forces us to choose the
high value $\chi_{\rm sel}=1.392$, for which we expect 4 foreground
stars to be included after cut 6, and we find a total of 12. After
applying also the colour cut 7, the number of stream candidate stars is
reduced to 4, with 1 expected to be foreground. It is not possible to
find more reliable stream candidates in this zone because of the high
value of $\chi_{\rm sel}$ we need to impose, but this situation will
improve in the future as the \textit{Gaia} proper motion errors and stream
model accuracy are improved.
\item[(iii)] \textit{Halo}: Here we can choose a much lower value of the
threshold, $\chi_{\rm sel}=5.65\pd{-3}$, with a corrected GOG18
expectation of 33 foreground stars after cut 6, and we find a total of
126. In this case, the foreground stars are much more effectively
eliminated by our cut 7, so only two foreground stars are expected after
cut 7 for the corrected GOG18 simulation. In contrast, in the GDR2
data, we find that 98 out of the 126 stars also pass cut 7, providing
strong evidence that these stars are indeed members of the M68 stream
that are near the distance inferred from our stream model. We also
remark that the 17 stars with the highest intersection with the stream
model (highest value of $\chi_{\rm sel}$) are all colour-magnitude
compatible with M68 and that 90 per cent are compatible among the 50 stars
with the highest $\chi_{\rm sel}$. We therefore expect most of the
98 stars in our final selection from this region to be true stream
members.
\item[(iv)] \textit{Disc 2}: In this region we actually obtain a larger
number of stars selected from GOG18 than from GDR2 when choosing a high
value of $\chi_{\rm sel}$, probably due to a Poisson fluctuation. This
indicates that the number of stream members in this region is likely
to be already very low. We have not selected any stars from this region, although future improved
\textit{Gaia} data may allow interesting stream candidates to be identified.
\end{enumerate}
 With the first 3 zones together, we finally have 151 stars that are
compatible with the phase-space model of the stellar stream (cut 6), out
of which 115 are also compatible with the HR-diagram of M68 (cut 7). If
our estimate of foreground contamination from GOG18 is correct, we
expect an average of only 4 of the final 115 stellar stream candidates
to be chance foreground projections. These stars are plotted in the
bottom panel of Fig. \ref{ad_sel} as large dots, with the red ones
being the stars that pass the cut 7 as well.

\begin{figure}
\begin{tabular}{c}
\includegraphics[width=0.95\columnwidth]{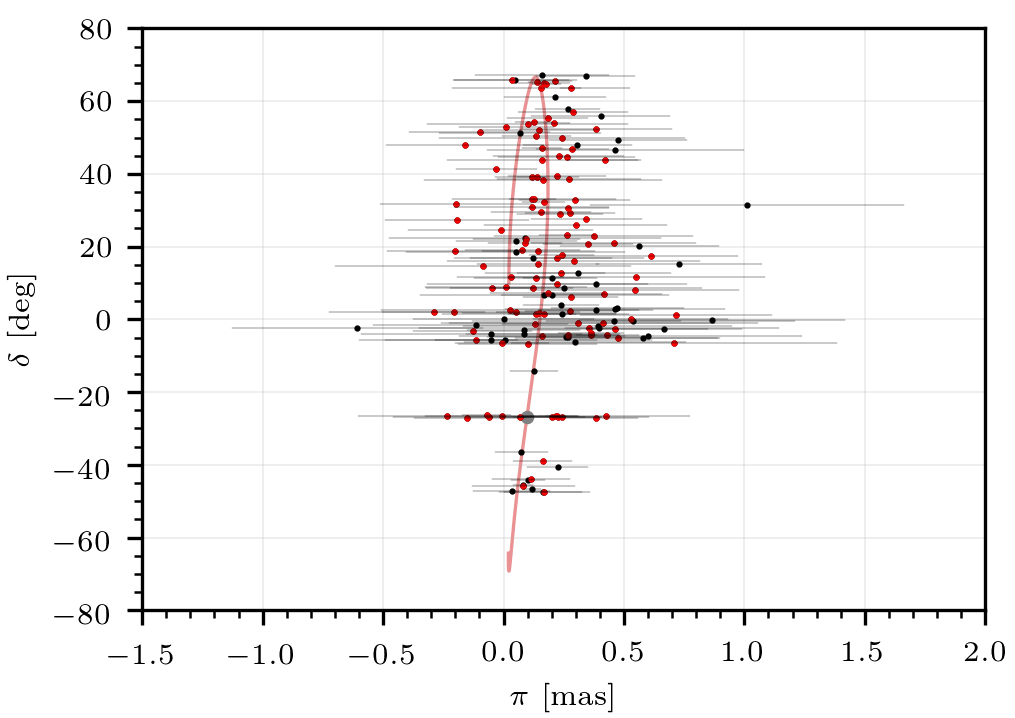}\\
\includegraphics[width=0.95\columnwidth]{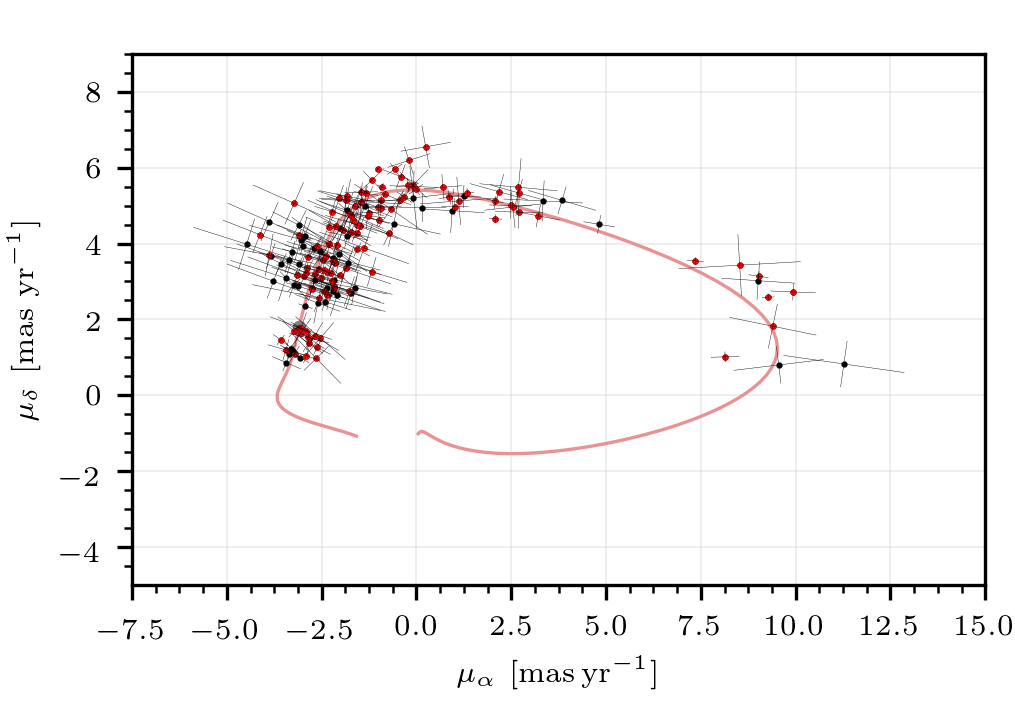}\\[0.5em]
\includegraphics[width=0.95\columnwidth]{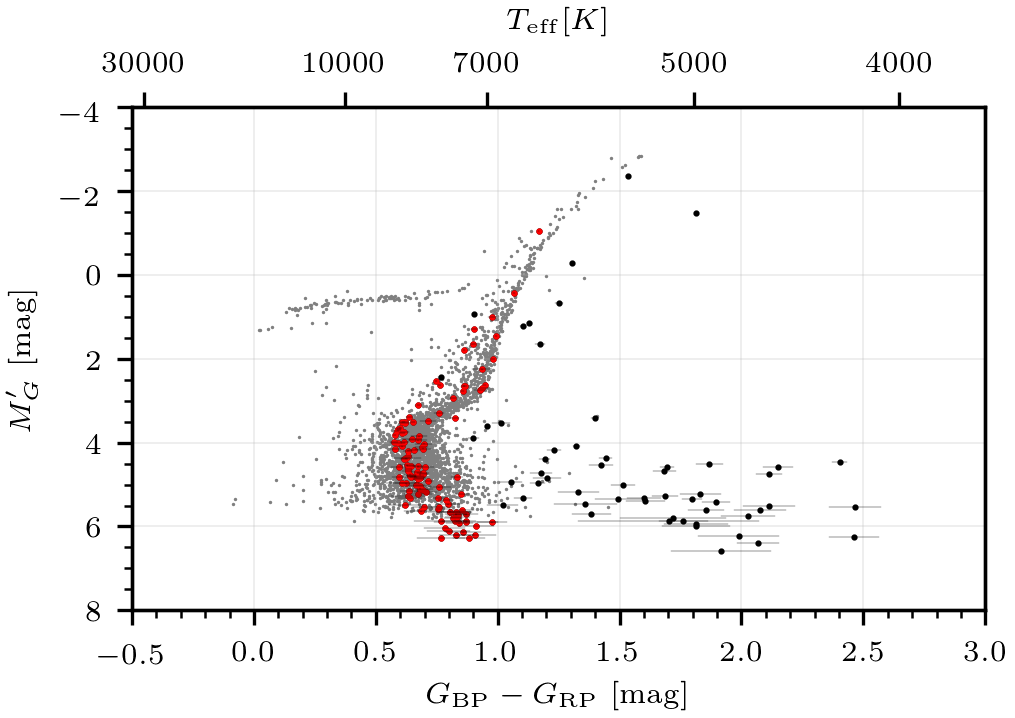}\\
\end{tabular}
\caption{Distributions of the finally selected 151 stars after our cut
6 using the kinematic data. Red dots show the 115 stars that pass also
our cut 7 of compatibility with the M68 HR-diagram and black dots are
the other 36 stars. \textit{Top}: Declination versus parallax.
\textit{Middle}: Proper motions. Grey dot marks the current position of
M68 and red line the orbit of the cluster during $100$ Myr backwards
and forwards from its current position. \textit{Bottom}: Grey dots are
stars that have been used to make the magnitude-space density model of
M68. Red dots have been selected to be compatible, whereas black dots
are not, using the inferred distance from the stream model.}
\label{ps_sel}
\end{figure}

  These finally selected stars are also shown in three panels in Fig.
\ref{ps_sel}, where we see their distribution in parallax, proper
motions and in the M68 HR-diagram. The red curve in the first two panels
is the expected trajectory from our stream model. Parallaxes are mostly
of insufficient accuracy to test the predicted distance to these stars,
and were mostly used in the pre-selection stage to rule out nearby
foreground disc stars. The middle panel, showing proper motions,
reflects also the consistency with the stream model. The HR-diagram in
the bottom panel shows that most of the final stream candidates are
inferred to be near the main-sequence turn-off. 

A list of the 115 stars passing our final selection is included in
Appendix \ref{App8}. GDR2 does not provide a radial velocity for any of these
stars. We have also checked the RAVE DR5 \citep{2017AJ....153...75K} and the LAMOST DR4 \citep{2015RAA....15.1095L}
catalogues, and have not found any of these stars.

\section{Conclusions}\label{con}

 A new method is presented in this paper to search for tidal streams,
based on maximization of a likelihood function that is calculated from a
model of the stream and of the foreground stellar population. The method
identifies the stream when there is sufficient statistical evidence that
a subset of the stars in the given catalogue are compatible with a stream
generated from a progenitor orbit that is fitted in the maximization
procedure, together with parameters of the gravitational potential. The
stream can be modelled as a superposition of Gaussians, which facilitates
the inclusion of the intrinsic stream dispersion and observational errors
for computing a realistic likelihood function. We present tests of the
method, and its first application to the tidal stream of the globular
cluster M68.

The stream we find coincides with the one previously discovered by
\citet{2019ApJ...872..152I}, who named it Fj\"orm. They found the stream
in a blind search using the Streamfinder method
\citep{2018MNRAS.477.4063M}, based on using six-dimensional tubes in
phase-space with the expected stream dispersion and counting the number
of stars compatible with a single stream. We have instead identified
this stream by searching specifically for one associated with M68. The
stream is detected in our method as an overdensity with respect to a
phase-space model of the Milky Way that is fitted to a physical
simulation of the stream caused by tidal shocking of the globular
cluster, by adjusting the Galactic gravitational potential and the
orbit of M68.

 The resulting orbit of our fit is fully consistent with the measured
kinematics of M68 and a simple Milky Way potential with only four free
parameters. Although we find preliminary constraints on the potential in
this paper, this needs to be further explored in future work, combining
with data from other known streams and allowing for realistic variations
of the potential contributed by the halo, disc and bulge that are
compatible with other observations. So far, we have discovered one of
the most visually obvious and easily detectable streams. As our method
is improved with an increasingly accurate foreground model and \textit{Gaia}
selection function and a more flexible and realistic parameterization
of the Milky Way gravitational potential, many more streams containing
fewer stars that are not obvious to the eye should likely be discovered.

 The M68 stream we have found is particularly promising to constrain
the Milky Way potential and to study the dynamical process of mass
segregation and tidal perturbations on a globular cluster as it crosses
the disc. The stream passes within only 5 kpc from us, implying that
accurate velocities from proper motions of relatively low-luminosity
stars are easier to measure than for other streams. Accurate radial
velocities of the stars we identify as likely stream members will also
add new constraints, and
surveys reaching down to fainter magnitudes over the sky region of the
stream should measure the abundance of low-mass main-sequence stars
that are below the \textit{Gaia} flux limit of detection. There is also a
promising potential to study the variable rate at which stars in M68
are inserted into the tidal stream from the distribution of velocities
in the stream stars, which should reflect the peaks of insertion
associated with disc crossings.

  In summary, the discovery of the M68 tidal stream opens the way to
increasing the sample of tidal streams and using them to determine
the potential of the Milky Way and to study the physical process of
tidal perturbation of clusters orbiting the Milky Way and creation of
the streams. In the future, we foresee the detection of new tidal streams by systematically applying our statistical method to other globular clusters with similar characteristics than M68.


\section*{Acknowledgements}
We would like to thank Mark Gieles and Daisuke Kawata for discussions, and
for pointing out the equivalence of the M68 tidal stream and the Fj\"orm
stream discovered by Ibata et al. We are grateful with Hamish Silverwood for his comments that helped us to improve this paper. We also acknowledge Francesca Figueras, Roger Mor and the rest of the \textit{Gaia} Team at University of Barcelona for useful discussions and helpful insights.
The authors have been supported by Spanish grant AYA2015-71091-P. Jordi Miralda-Escud\'e was also
supported by Spanish grant PRX18/00444 and by the Corning Glass Works
Foundation Fellowship Fund for a stay at Institute for Advanced Study (IAS), Princeton.
This work has made use of data from the European Space Agency (ESA) mission
\textit{Gaia} (\url{https://www.cosmos.esa.int/gaia}), processed by the \textit{Gaia} Data
Processing and Analysis Consortium
(DPAC, \url{https://www.cosmos.esa.int/web/gaia/dpac/consortium}). Funding
for the DPAC has been provided by national institutions, in particular, the
institutions participating in the \textit{Gaia} Multilateral Agreement.



\bibliographystyle{mnras}
\bibliography{bib/ref.bib}



\clearpage
\newpage 

\appendix

\newcounter{defC}
\setcounter{defC}{1}
\newcounter{prpC}
\setcounter{prpC}{1}


\section{Coordinate transformations}\label{App2}

 The phase-space coordinates of the stars are transformed to the
Galactocentric Coordinate System using the solar position and velocity
from \citet{2016ARA&A..54..529B}. The position of the Sun,
$x^i_{\odot}$, in the Galactocentric Coordinate System and Cartesian
coordinates $(x,y,z)$, is
\begin{equation}
x^i_{\odot} = \left(\!\!\!
 \begin{tabular}{c}$8.2\pm0.1$\\$0$\\$0.025\pm0.005$\end{tabular}\!\!\!
 \right) \quad\units{kpc} ~.
\end{equation}

 The velocity of the Sun, with the components with respect to the LSR
$U$, pointing to the Galactic Centre,
$V$, positive along the direction of the Sun's
rotation, and $W$, positive towards the North Galactic
Pole, are:
\begin{equation}
v_\odot \equiv \left(\!\!\!
 \begin{tabular}{c}$v_{\scaleto{U}{3.5pt}}$\\
  $v_{\scaleto{V}{3.5pt}}$\\$v_{\scaleto{W}{3.5pt}}$\end{tabular}\!\!\!
\right) = \left(\!\!\!
 \begin{tabular}{c}14\\12.24\\7.25\end{tabular}\!\!\!\right)
 \quad\units{km s$^{-1}$} ~.
\end{equation}
The rotational velocity of the LSR along $\scaleto{V}{4.5pt}$ is assumed to be
\begin{equation}
v_\SM{LSR}= 238\pm15 \quad\units{km s$^{-1}$} ~.
\end{equation}


\section{Colour-Magnitude diagram of M68}\label{App4}

\noindent We reproduce here the ADQL query we have used to obtain the
GDR2 photometry in $G$, $G_{\rm BP}$, and
$G_{\rm RP}$ passbands in a circle centred on M68, giving
2929 stars:

\begin{lstlisting}[language=SQL, showspaces=false, basicstyle=\ttfamily, numbers=left, numberstyle=\tiny, commentstyle=\color{gray}, breaklines=true, keywords={AS, SELECT, FROM, WHERE, BETWEEN, IS, NOT, NULL, AND, CONTAINS, POINT, CIRCLE, COUNT}, keywordstyle=\color{mymauve}, stringstyle=\color{blue}, commentstyle=\color{mygreen}, xleftmargin=4.6mm]
SELECT bp_rp, phot_g_mean_mag, phot_bp_mean_flux, phot_bp_mean_flux_error, phot_rp_mean_flux, phot_rp_mean_flux_error, phot_g_mean_flux, phot_g_mean_flux_error 
FROM gdr2.gaia_source 
WHERE 1 = CONTAINS( POINT('ICRS', ra, dec), CIRCLE('ICRS', 189.8651, -26.7454, 0.1) ) 
AND parallax <= 10.0 
AND SQRT((pmra+2.78)*(pmra+2.78) + (pmdec-1.81)*(pmdec-1.81)) <= 1.78 
AND bp_rp <= 2.0;
\end{lstlisting}

\begin{flushleft}
\phantom{a}
Host server: \url{https://gaia.aip.de/}\\
Description of the \texttt{gaia\_source} table: \url{https://gea.esac.esa.int/archive/documentation/GDR2/Gaia_archive/chap_datamodel/sec_dm_main_tables/ssec_dm_gaia_source.html}\\
\end{flushleft}


\section{The Pre-selection}

\subsection{General method}\label{App6_met}

 Given a tidal stream progenitor, a set of $M$ orbits with phase-space
components $\eta_m^\mu\var{t}$ (where $m=1,\ldots,M$) are computed over
a time interval $-l < t < l$, from the present phase-space coordinates
$\eta^\mu_m\var{0}$ and different values of the potential free
parameters $\theta_{\phi m}$, following the distributions:
\begin{align}
&\eta^\mu_m\var{0} \sim G\vvar{\bar{\eta}^\mu,\varepsilon_\mu^2} ~, \\[0.5em]
&\theta_{\phi m} \sim U\vvar{\bar{\theta}_{\phi}-\varepsilon_\phi,
 \bar{\theta}_{\phi}+\varepsilon_\phi} ~,
\end{align}
where $G$ is a Gaussian distribution and $U\var{b,c}$ is a uniform
distribution over $b < x < c$.

 The bundle of $M$ orbits defines a phase-space region with a
probability density $V$ of finding stars belonging to any orbit in the
bundle. We smooth the distribution of the simulated orbits describing
it as the sum of $N-1$ Gaussian distributions, obtained from $N$ points
along each orbit uniformly distributed in time, labelled by the index
$n=0,\ldots,N$ with time intervals $\Delta t\equiv 2l/N$, and defining
\begin{equation}
\eta^\mu_{mn} \equiv \eta^\mu_m\var{-l+n\Delta t} ~.
\end{equation}
The means and covariance matrices of the Gaussian distributions are
computed as:
\begin{equation}
\bar{\eta}^\mu_n = \frac{1}{M} \sum_{m=1}^M \eta^\mu_{mn} \,
 \qquad\qquad 0<n<N ~;
\end{equation}
\begin{equation}
\Xi^{\mu\nu}_n = \frac{1}{3M}\!
 \sum_{i=n-1}^{n+1} \sum_{m=1}^{M} (\eta^\mu_{mi} - \bar{\eta}^\mu_n)
 (\eta^\nu_{mi} - \bar{\eta}^\nu_n) \quad\quad 0<n<N \, .
\end{equation}
The distribution is given by
\begin{equation}
V\var{w^\mu} \equiv \frac{1}{N-1} \sum_{n=1}^{N-1} G\vvar{w^\mu-\bar{\eta}^\mu_n|\Xi^{\mu\nu}_n} ~.
\end{equation}

 The intersection of a star with observed phase-space coordinates
$w^\mu_o$ and errors $\sigma^{\mu\nu}$ with the region $V$
is now expressed as the convolution of the two Gaussian distributions,
\begin{equation}
P_\SM{REG} = \frac{1}{N-1}
 \sum_{n=1}^{N-1}
 G\vvar{w^\mu_o-\bar{\eta}^\mu_n|\sigma^{\mu\nu}+\Xi^{\mu\nu}_n} ~.
\end{equation}

\subsection{Pre-selection for M68}\label{App6_m68}

 For the case of M68, the region $V$ has been described with $N=101$
Gaussian distributions computed using a bundle of $M=100$ orbits of
length $l = 50$ Myr. The parameters used to compute the bundle of orbits
are listed in Table \ref{pre_par}. Stars obeying
$P_\SM{REG} \geqslant 1.4893\pd{-4}$ ${\rm yr}^3 \, {\rm deg}^{-2} \, {\rm pc}^{-1} \, {\rm mas}^{-3}$ have been chosen for our pre-selection.

 Table \ref{gc_pre} lists the globular clusters that lie inside the
pre-selection region and the angular radius $\xi$ of the circle within
which stars are removed.

\begin{table}
\caption[]{\small{
Pre-selection parameters used to compute the bundle of orbits for M68.}}

\begin{center}
\setlength{\tabcolsep}{3.5pt}
\begin{tabular}{lcccccc}
\toprule
&$\pi$&$\delta$&$\alpha$&$v_r$&$\mu_\delta$&$\mu_\alpha$\\
&\units{mas}&\units{deg}&\units{deg}&\units{km s$^{-1}$}&\units{mas yr$^{-1}$}&\units{mas yr$^{-1}$}\\
\midrule
$\bar{\eta}^\mu$&0.0971&-26.75&189.87&-94.7&1.7916&-3.0951\\
$\varepsilon_\mu$&0.0023&2.5&2.5&0.2&0.0039&0.0056\\
\midrule
&$\rho_{0\rm dh}$&$a_{1 \rm dh}$&$a_{3 \rm dh}$&$\beta_{\rm dh}$&&\\
&\units{M$_{\odot}$ kpc$^{-3}$}&\units{kpc}&\units{kpc}&&&\\
\midrule
$\bar{\theta}_\phi$&$8\pd{6}$&$20.2$&$16.16$&$3.1$&&\\
$\varepsilon_\phi$&$1\pd{6}$&$4$&$4$&$0.2$&&\\
\bottomrule
\end{tabular}
\end{center}

\label{pre_par}
\end{table}

\begin{table}
\caption[]{\small{Coordinates of the globular clusters that lie in the pre-selection region and radius of the angular circle.}}
\begin{center}
\begin{tabular}{llrrc}
\toprule
\multicolumn{2}{l}{\multirow{2}{*}{\textbf{Globular Cluster}}}&\multicolumn{1}{c}{$\delta$}&\multicolumn{1}{c}{$\alpha$}&$\xi$\\
&&\multicolumn{1}{c}{\units{deg}}&\multicolumn{1}{c}{\units{deg}}&\units{deg}\\
\midrule
&NGC5466&28.5331&211.3614&0.08\\
M3&NGC5272&28.3760&205.5486&0.2\\
M53&NGC5024&18.1661&198.2262&0.2\\
&NGC5053&17.7008&199.1124&0.2\\
M68&NGC4590&-26.7454&189.8651&0.3\\
\bottomrule
\end{tabular}
\end{center}

\begin{tabular}{l}
\textit{Note.} Ref.: \citet{2018AA...616A..12G}\\
\end{tabular}

\label{gc_pre}
\end{table}


\section{Colour-Magnitude Selection}\label{App7}

\subsection{Method}

 A colour-magnitude index of the progenitor cluster is first constructed
from a sample of $N_e$ stars with observed $\BPRP$ colour index and a
\Gband absolute magnitude $M_{G}'$. Defining a position of the $i$
star as $x^\mu_i \equiv (\BPRP,\, M_G')$, the density is modelled using
a Kernel Density Estimator with a Gaussian kernel, and with covariance
matrices
\begin{equation}
\varXi_i^{\mu\nu} \equiv
 \left(\sum_{j=1}^{N_e} c_{ij} \right)^{-1} \,\,
 \sum_{j=1}^{N_e} c_{ij} \,
 (x^\mu_j - x^\mu_i)\,(x^\nu_j - x^\nu_i) ~.
\end{equation}
Weights are defined using the constant $d_0=0.07$ mag,
\begin{equation}
c_{ij} \equiv \left(d_0+d_{ij}\,\right)^{5}  ~,
\qquad\quad d_{ij}^2\equiv\sum_{l=1}^{2}(x^l_j-x^l_i)^2  ~.
\end{equation}
Given the integrated $G$ mean flux $f_G$, its observational error $\varepsilon_{f_G}$ and assuming a symmetric error distribution and neglecting the uncertainty of the zero-point magnitude in the Vega scale, the error of the associated \Gband magnitude is
\begin{equation}\label{col}
\varepsilon_{G} = \frac{2.5}{\ln\!\var{10}}\frac{\varepsilon_{f_G}}{f_G} ~.
\end{equation}
The deviation of the colour index is computed as of equation (\ref{col}) for both magnitudes and subtracting their errors
\begin{equation}
\varepsilon_{BP-RP} = \sqrt{\varepsilon_{BP}^2-\varepsilon_{RP}^2} ~.
\end{equation}
\noindent In the case of GOG18, it is necessary to correct the discrepancy between the simulation and GDR2 catalogue. Defining the scale factor $\lambda_\mu\equiv(2, 1.6)$, the scaled errors are $\lambda_\mu\varepsilon_\mu$.

A star in magnitude-space is represented by a Gaussian distribution with mean observed position $x^\mu_o$ convolved with uncorrelated uncertainties, its covariance matrix is
\begin{equation*}
\sigma^{\mu\nu} \equiv \left\{ \!\!\!
  \begin{tabular}{ll}
  $\varepsilon^2_\mu $ & $\mu=\nu$\\
   &   \\[-1ex]
  $0$ & $\mu \neq \nu$
  \end{tabular}
\right. ~.
\end{equation*}
\noindent Neglecting dust extinction, the intersection between a star
and the density model is given by the convolution
\begin{equation}
P_\SM{CR} = \frac{1}{N_e} \sum_{i=1}^{N_e} G\vvar{x_o^\mu - x^\mu_i | \sigma^{\mu\nu}+\Xi^{\mu\nu}_i} ~.
\end{equation}

\subsection{Colour-Magnitude selection for M68}

For M68, we use a sample of $N_e=2929$ stars to construct the density
model, using the selection described in Appendix \ref{App4}. Stars with
$P_\SM{CR} \geqslant 0.08$ mag$^{-2}$ have been selected. 


\section{Selected Stars}\label{App8}

The selected stars from GDR2 catalogue are listed in Table \ref{sel0}.

\begin{table*}
\caption[]{\small{Selected stars from GDR2 catalogue. They are compatible with a phase-space density model of the tidal stream of M68 and with its HR-diagram. GDR2 does not provide radial velocity for any of these stars.}}
\begin{center}
\begin{tabular}{rrrrrrrrrr}
\toprule

\multicolumn{1}{c}{N}
&\multicolumn{1}{c}{source\_id}
&\multicolumn{1}{c}{$\pi$}
&\multicolumn{1}{c}{$\delta$}
&\multicolumn{1}{c}{$\alpha$}
&\multicolumn{1}{c}{$\mu_\delta$}
&\multicolumn{1}{c}{$\mu_{\alpha*}$}
&\multicolumn{1}{c}{\scalebox{0.8}{$G_{\rm BP}\!-\!G_{\rm RP}$}}
&\multicolumn{1}{c}{$G$}
&\multicolumn{1}{c}{$\chi_{\rm sel}$}\\
&
&\multicolumn{1}{c}{\units{mas}}
&\multicolumn{1}{c}{\units{deg}}
&\multicolumn{1}{c}{\units{deg}}
&\multicolumn{1}{c}{\units{mas yr$^{-1}$}}
&\multicolumn{1}{c}{\units{mas yr$^{-1}$}}
&\multicolumn{1}{c}{\units{mag}}
&\multicolumn{1}{c}{\units{mag}}
&\multicolumn{1}{c}{\units{\scalebox{0.8}{yr$^{3}$ deg$^{-2}$ pc$^{-1}$ mas$^{-3}$}}}\\

\midrule

1&3496364826490984832&$0.0666$&$-26.9401$&$189.5268$&$1.7746$&$-2.7097$&$1.1689$&$14.0009$&$7.9446\text{E}\:\!\:\!\plus\:\!02$\\
2&3496397262084464128&$0.2211$&$-26.6147$&$190.1435$&$1.7706$&$-2.7786$&$0.9929$&$16.4954$&$3.8387\text{E}\:\!\:\!\plus\:\!01$\\
3&6133483847268997632&$0.1142$&$-44.0194$&$190.4956$&$1.0431$&$-2.0908$&$0.8980$&$17.6564$&$2.2383\text{E}\:\!\:\!\plus\:\!00$\\
4&3496359908751562496&$-0.1533$&$-27.0259$&$189.6791$&$1.5048$&$-2.5199$&$0.7139$&$18.5297$&$1.8531\text{E}\:\!\:\!\plus\:\!00$\\
5&6129336321904932224&$0.0825$&$-45.8545$&$191.0678$&$1.1946$&$-2.3981$&$0.9346$&$18.3756$&$1.6943\text{E}\:\!\:\!\plus\:\!00$\\
6&3496403270742208768&$-0.0047$&$-26.5809$&$190.1661$&$1.5555$&$-2.4028$&$0.6775$&$18.9026$&$1.6537\text{E}\:\!\:\!\plus\:\!00$\\
7&3496413819180463616&$0.2016$&$-26.8035$&$189.4730$&$1.3666$&$-2.5346$&$0.7607$&$18.3398$&$1.6298\text{E}\:\!\:\!\plus\:\!00$\\
8&6128205985298801408&$0.1692$&$-47.5466$&$191.4542$&$1.0899$&$-2.1682$&$0.9766$&$17.2226$&$1.5785\text{E}\:\!\:\!\plus\:\!00$\\
9&6153341199065073536&$0.1632$&$-38.9526$&$189.3009$&$1.4558$&$-2.7846$&$0.9022$&$16.9682$&$1.4216\text{E}\:\!\:\!\plus\:\!00$\\
10&3496363825761763840&$0.2272$&$-26.9772$&$189.5723$&$1.7695$&$-2.7182$&$0.6242$&$19.4263$&$9.9037\text{E}\:\!\:\!\minus\:\!01$\\
11&3496426983257641216&$0.2179$&$-26.5627$&$189.5454$&$1.6971$&$-2.5965$&$1.0683$&$15.4868$&$9.6922\text{E}\:\!\:\!\minus\:\!01$\\
12&3496351250097506816&$0.3829$&$-27.0117$&$190.0502$&$1.6874$&$-2.8738$&$0.9253$&$17.7917$&$9.5428\text{E}\:\!\:\!\minus\:\!01$\\
13&3496465702385628928&$-0.0691$&$-26.3824$&$189.6310$&$1.5405$&$-2.5617$&$0.9822$&$17.0284$&$9.2144\text{E}\:\!\:\!\minus\:\!01$\\
14&3496383857489983232&$0.4248$&$-26.6971$&$190.2375$&$1.6487$&$-2.7174$&$0.6725$&$18.9987$&$8.5381\text{E}\:\!\:\!\minus\:\!01$\\
15&3496417912286695936&$0.2415$&$-26.7512$&$189.3964$&$1.2602$&$-2.3346$&$0.9350$&$17.7570$&$7.2085\text{E}\:\!\:\!\minus\:\!01$\\
16&3496385266239278976&$-0.2349$&$-26.6352$&$190.1922$&$1.5076$&$-2.2656$&$0.6959$&$19.0733$&$6.7434\text{E}\:\!\:\!\minus\:\!01$\\
17&3692120708467214208&$0.0521$&$+2.0591$&$194.8744$&$3.2952$&$-2.4461$&$0.8598$&$16.7081$&$5.9697\text{E}\:\!\:\!\minus\:\!01$\\
18&3496354101955858432&$-0.0606$&$-26.9265$&$190.1141$&$0.9713$&$-2.3555$&$0.6465$&$18.9647$&$5.5831\text{E}\:\!\:\!\minus\:\!01$\\
19&1606236095606481280&$0.1245$&$+54.2566$&$223.1041$&$5.1230$&$1.2085$&$0.5928$&$17.4589$&$2.5816\text{E}\:\!\:\!\minus\:\!01$\\
20&1458389959637031296&$0.1198$&$+33.0455$&$206.2680$&$5.4940$&$-0.7652$&$0.5769$&$17.4817$&$1.9672\text{E}\:\!\:\!\minus\:\!01$\\
21&3677284104720388224&$-0.0056$&$-6.5245$&$193.5543$&$2.6362$&$-2.3346$&$0.6176$&$17.7979$&$1.5302\text{E}\:\!\:\!\minus\:\!01$\\
22&3730942295084892928&$-0.0484$&$+8.7147$&$197.5589$&$3.9250$&$-2.5897$&$0.6311$&$18.0954$&$1.2550\text{E}\:\!\:\!\minus\:\!01$\\
23&1455844345403174016&$0.2761$&$+29.3251$&$204.2349$&$5.3560$&$-1.2723$&$0.6308$&$18.3614$&$1.1934\text{E}\:\!\:\!\minus\:\!01$\\
24&3691856546503440128&$0.1471$&$+1.7067$&$195.5511$&$3.4746$&$-2.1421$&$0.6608$&$18.8550$&$1.1440\text{E}\:\!\:\!\minus\:\!01$\\
25&3675964587688060416&$0.1004$&$-6.6935$&$192.9816$&$2.7575$&$-2.4202$&$0.6885$&$18.3793$&$1.1291\text{E}\:\!\:\!\minus\:\!01$\\
26&1456842495802767744&$0.1670$&$+32.3278$&$205.7151$&$5.3068$&$-0.7095$&$0.5872$&$17.3960$&$1.0621\text{E}\:\!\:\!\minus\:\!01$\\
27&3678259474613592960&$0.3630$&$-4.1679$&$195.2089$&$3.1499$&$-2.9547$&$0.8179$&$17.1344$&$1.0476\text{E}\:\!\:\!\minus\:\!01$\\
28&3729366935440429952&$0.2786$&$+6.2290$&$195.7788$&$3.5696$&$-2.4608$&$0.6158$&$17.9548$&$1.0185\text{E}\:\!\:\!\minus\:\!01$\\
29&3678958794072858112&$0.4306$&$-4.2433$&$193.8940$&$3.0850$&$-2.5129$&$0.6187$&$17.7100$&$9.4803\text{E}\:\!\:\!\minus\:\!02$\\
30&3938507382917888512&$0.2424$&$+17.7672$&$199.7684$&$4.2657$&$-1.4890$&$0.6077$&$17.5658$&$7.7959\text{E}\:\!\:\!\minus\:\!02$\\
31&3730306639925232384&$0.1831$&$+7.2101$&$196.3747$&$3.9700$&$-2.0812$&$0.6286$&$18.4781$&$7.5430\text{E}\:\!\:\!\minus\:\!02$\\
32&3736013929907004928&$0.1335$&$+11.2735$&$197.5279$&$4.4459$&$-2.2741$&$0.6227$&$18.8557$&$6.9948\text{E}\:\!\:\!\minus\:\!02$\\
33&1504696231140527232&$0.4210$&$+43.8973$&$212.8308$&$5.5413$&$-0.1608$&$0.6686$&$18.3991$&$6.9580\text{E}\:\!\:\!\minus\:\!02$\\
34&3686001101624868352&$0.1303$&$-1.1354$&$195.9989$&$3.2547$&$-2.3681$&$0.6370$&$17.5221$&$6.0207\text{E}\:\!\:\!\minus\:\!02$\\
35&1605773136787896320&$0.0124$&$+52.9402$&$221.9906$&$5.0068$&$1.4966$&$0.6926$&$18.9247$&$4.8143\text{E}\:\!\:\!\minus\:\!02$\\
36&3731064894927486976&$0.1237$&$+8.6340$&$196.1463$&$4.2136$&$-3.0731$&$0.5780$&$18.0480$&$4.5961\text{E}\:\!\:\!\minus\:\!02$\\
37&1455852248142764288&$0.1545$&$+29.4651$&$204.2691$&$5.1368$&$-0.8114$&$0.6936$&$18.4129$&$4.5207\text{E}\:\!\:\!\minus\:\!02$\\
38&3677590211334628992&$-0.1124$&$-5.6794$&$193.4493$&$2.8830$&$-2.1745$&$0.6168$&$18.0113$&$4.4703\text{E}\:\!\:\!\minus\:\!02$\\
39&3677713562794613504&$0.4733$&$-5.2572$&$193.5804$&$2.8342$&$-2.1752$&$0.7574$&$19.3068$&$4.3766\text{E}\:\!\:\!\minus\:\!02$\\
40&1600425932567734784&$0.1829$&$+55.3254$&$226.6320$&$4.7296$&$1.8281$&$0.7136$&$18.7518$&$4.2661\text{E}\:\!\:\!\minus\:\!02$\\
41&1603928701735870336&$0.1355$&$+50.4873$&$219.6809$&$4.9731$&$0.6445$&$0.6951$&$18.5037$&$4.2338\text{E}\:\!\:\!\minus\:\!02$\\
42&3736784721917729152&$0.2377$&$+12.7628$&$198.6407$&$3.8432$&$-1.5178$&$0.6232$&$18.2278$&$4.0618\text{E}\:\!\:\!\minus\:\!02$\\
43&3939089398230332928&$0.0781$&$+19.1721$&$200.8586$&$4.4500$&$-1.3995$&$0.6745$&$18.6470$&$4.0158\text{E}\:\!\:\!\minus\:\!02$\\
44&1634945752956687104&$0.1780$&$+64.8292$&$251.4316$&$3.1381$&$3.8373$&$0.8572$&$16.9145$&$3.8506\text{E}\:\!\:\!\minus\:\!02$\\
45&1505103084803276160&$0.2284$&$+44.9951$&$213.9660$&$5.4822$&$0.4941$&$0.7598$&$19.2384$&$3.5483\text{E}\:\!\:\!\minus\:\!02$\\
46&3736463664522214656&$0.0303$&$+11.7794$&$198.7358$&$4.2984$&$-1.6778$&$0.6700$&$18.6768$&$3.4229\text{E}\:\!\:\!\minus\:\!02$\\
47&1506844779940816384&$0.2853$&$+46.7914$&$214.8890$&$5.2314$&$0.5913$&$0.6458$&$18.2705$&$3.4038\text{E}\:\!\:\!\minus\:\!02$\\
48&3729915076346434304&$0.4170$&$+7.1123$&$197.1414$&$3.5768$&$-2.1621$&$0.7025$&$18.5110$&$3.3254\text{E}\:\!\:\!\minus\:\!02$\\
49&3678969243729036032&$0.2682$&$-4.1958$&$193.5954$&$3.3351$&$-2.5662$&$0.9471$&$16.8203$&$3.2393\text{E}\:\!\:\!\minus\:\!02$\\
50&3690547165593837312&$0.1666$&$+1.4787$&$193.0467$&$3.3546$&$-2.8934$&$0.6077$&$17.5944$&$3.1051\text{E}\:\!\:\!\minus\:\!02$\\
51&1496589879802806528&$0.1163$&$+39.0498$&$207.1911$&$5.9647$&$-0.7894$&$0.6743$&$16.7684$&$3.0905\text{E}\:\!\:\!\minus\:\!02$\\
52&1496266863902585728&$0.2228$&$+39.5427$&$209.3919$&$5.2231$&$-0.2448$&$0.5933$&$18.2511$&$2.9648\text{E}\:\!\:\!\minus\:\!02$\\
53&1496030911283208576&$0.1392$&$+39.1041$&$210.2471$&$5.4324$&$-0.0120$&$0.6456$&$18.4847$&$2.8611\text{E}\:\!\:\!\minus\:\!02$\\
54&1498889611451314816&$-0.0300$&$+41.4624$&$209.2182$&$5.9579$&$-0.4318$&$0.6810$&$18.5142$&$2.7590\text{E}\:\!\:\!\minus\:\!02$\\
55&1604273437287099776&$-0.0990$&$+51.4980$&$219.8012$&$5.3673$&$1.3525$&$0.8035$&$19.4194$&$2.7511\text{E}\:\!\:\!\minus\:\!02$\\
56&3678292459962501376&$0.3620$&$-3.8740$&$195.7269$&$3.2026$&$-2.6585$&$0.8230$&$17.6250$&$2.6342\text{E}\:\!\:\!\minus\:\!02$\\
57&3685425060611225472&$0.4647$&$-2.5588$&$194.5278$&$3.5328$&$-2.2110$&$0.6830$&$19.2702$&$2.6098\text{E}\:\!\:\!\minus\:\!02$\\

\bottomrule
\end{tabular}
\end{center}

\label{sel0}
\end{table*}

\begin{table*}\addtocounter{table}{-1}
\caption[]{\small{\textit{- continued}}}
\begin{center}
\begin{tabular}{rrrrrrrrrr}
\toprule

\multicolumn{1}{c}{N}
&\multicolumn{1}{c}{source\_id}
&\multicolumn{1}{c}{$\pi$}
&\multicolumn{1}{c}{$\delta$}
&\multicolumn{1}{c}{$\alpha$}
&\multicolumn{1}{c}{$\mu_\delta$}
&\multicolumn{1}{c}{$\mu_{\alpha*}$}
&\multicolumn{1}{c}{\scalebox{0.8}{$G_{\rm BP}\!-\!G_{\rm RP}$}}
&\multicolumn{1}{c}{$G$}
&\multicolumn{1}{c}{$\chi_{\rm sel}$}\\
&
&\multicolumn{1}{c}{\units{mas}}
&\multicolumn{1}{c}{\units{deg}}
&\multicolumn{1}{c}{\units{deg}}
&\multicolumn{1}{c}{\units{mas yr$^{-1}$}}
&\multicolumn{1}{c}{\units{mas yr$^{-1}$}}
&\multicolumn{1}{c}{\units{mag}}
&\multicolumn{1}{c}{\units{mag}}
&\multicolumn{1}{c}{\units{\scalebox{0.8}{yr$^{3}$ deg$^{-2}$ pc$^{-1}$ mas$^{-3}$}}}\\

\midrule

58&3731172780210011264&$0.0096$&$+8.9611$&$196.7009$&$3.6435$&$-2.8289$&$0.6399$&$19.2347$&$2.5511\text{E}\:\!\:\!\minus\:\!02$\\
59&3678889254257425664&$0.1613$&$-4.5256$&$194.3135$&$3.6767$&$-2.3650$&$0.6577$&$18.3852$&$2.4068\text{E}\:\!\:\!\minus\:\!02$\\
60&1605598138344043904&$0.1489$&$+52.1831$&$221.9488$&$4.8273$&$1.6527$&$0.8305$&$19.4252$&$2.3852\text{E}\:\!\:\!\minus\:\!02$\\
61&3690382170130121984&$0.1339$&$+1.3929$&$194.3795$&$3.2183$&$-2.9074$&$0.6628$&$19.0984$&$2.2914\text{E}\:\!\:\!\minus\:\!02$\\
62&1456411624683879296&$0.1173$&$+30.7712$&$205.5668$&$4.9705$&$-0.8631$&$0.6354$&$18.8292$&$2.2258\text{E}\:\!\:\!\minus\:\!02$\\
63&1442304443823114752&$0.4577$&$+21.0880$&$201.0665$&$5.2465$&$-1.7063$&$0.7897$&$19.1160$&$2.1508\text{E}\:\!\:\!\minus\:\!02$\\
64&3744325138301094784&$0.1423$&$+15.3030$&$200.4737$&$4.3564$&$-1.8696$&$0.8476$&$19.0354$&$2.0853\text{E}\:\!\:\!\minus\:\!02$\\
65&3688452737676950400&$0.3545$&$-2.2653$&$194.0000$&$3.0246$&$-2.2345$&$0.7619$&$16.7977$&$2.0055\text{E}\:\!\:\!\minus\:\!02$\\
66&1603533873982622592&$0.2436$&$+49.9313$&$219.0180$&$5.1240$&$0.7174$&$0.8574$&$19.8819$&$1.9860\text{E}\:\!\:\!\minus\:\!02$\\
67&3702709520838936064&$0.2749$&$+2.3905$&$194.0326$&$3.2250$&$-2.2645$&$0.5742$&$18.0258$&$1.9167\text{E}\:\!\:\!\minus\:\!02$\\
68&3940514399659927296&$0.0898$&$+20.9339$&$199.6473$&$4.4686$&$-1.9810$&$0.6312$&$18.0733$&$1.9045\text{E}\:\!\:\!\minus\:\!02$\\
69&1606249122242914688&$0.2103$&$+53.9305$&$224.3375$&$4.6520$&$1.2271$&$0.6511$&$17.3305$&$1.6830\text{E}\:\!\:\!\minus\:\!02$\\
70&1629009390895385728&$0.2782$&$+63.7497$&$246.4188$&$3.5416$&$3.2517$&$0.7459$&$16.6089$&$1.5032\text{E}\:\!\:\!\minus\:\!02$\\
71&1448109899577010688&$0.2991$&$+25.9842$&$203.5458$&$5.1577$&$-1.6772$&$0.8423$&$19.6347$&$1.4980\text{E}\:\!\:\!\minus\:\!02$\\
72&3690318879491794816&$0.7154$&$+1.0770$&$194.9856$&$3.2560$&$-1.1810$&$0.6733$&$19.3076$&$1.4842\text{E}\:\!\:\!\minus\:\!02$\\
73&3692071982062671872&$0.0273$&$+2.6639$&$195.8328$&$3.9871$&$-2.3096$&$0.6086$&$19.0058$&$1.4675\text{E}\:\!\:\!\minus\:\!02$\\
74&1504902668745341568&$0.2623$&$+44.7623$&$214.4142$&$6.1911$&$-0.1412$&$0.7564$&$19.2867$&$1.4605\text{E}\:\!\:\!\minus\:\!02$\\
75&3691940624783331968&$-0.2046$&$+1.9962$&$195.6238$&$3.6968$&$-3.8815$&$0.7665$&$19.9206$&$1.4019\text{E}\:\!\:\!\minus\:\!02$\\
76&1606240768531075840&$0.0999$&$+53.7979$&$224.1231$&$5.4837$&$1.5892$&$0.7981$&$19.9199$&$1.3801\text{E}\:\!\:\!\minus\:\!02$\\
77&1635385046508786432&$0.1401$&$+65.3460$&$254.9531$&$2.5866$&$3.8704$&$0.8620$&$15.9535$&$1.3366\text{E}\:\!\:\!\minus\:\!02$\\
78&3744739963422459648&$0.2229$&$+16.8007$&$199.5876$&$4.9930$&$-1.5486$&$0.8518$&$19.3988$&$1.3101\text{E}\:\!\:\!\minus\:\!02$\\
79&3939039542249768704&$-0.2004$&$+18.7968$&$200.7388$&$5.1992$&$-1.9285$&$0.7052$&$18.9435$&$1.2349\text{E}\:\!\:\!\minus\:\!02$\\
80&1602451748381843840&$0.2900$&$+57.0243$&$230.1132$&$5.3363$&$1.4649$&$0.6810$&$18.9183$&$1.2198\text{E}\:\!\:\!\minus\:\!02$\\
81&1454872239685951232&$0.2356$&$+28.8703$&$206.4059$&$5.3194$&$-1.1628$&$0.6408$&$18.4768$&$1.1805\text{E}\:\!\:\!\minus\:\!02$\\
82&3688651916785064320&$0.4148$&$-1.1125$&$195.0655$&$3.1597$&$-3.1376$&$0.8037$&$19.7906$&$1.1723\text{E}\:\!\:\!\minus\:\!02$\\
83&1448864714313741184&$0.3428$&$+27.5361$&$203.2331$&$4.9445$&$-0.8370$&$0.6315$&$18.3315$&$1.1234\text{E}\:\!\:\!\minus\:\!02$\\
84&2257502327664371328&$0.2131$&$+65.4879$&$273.9369$&$1.0117$&$3.3762$&$0.8663$&$17.0513$&$1.1005\text{E}\:\!\:\!\minus\:\!02$\\
85&1495975420305337728&$0.2719$&$+38.5514$&$209.8212$&$5.6724$&$-0.9194$&$0.8382$&$19.4064$&$1.0926\text{E}\:\!\:\!\minus\:\!02$\\
86&1506911987588666112&$0.1596$&$+47.0514$&$215.9830$&$5.3300$&$0.9153$&$0.5956$&$17.4090$&$1.0210\text{E}\:\!\:\!\minus\:\!02$\\
87&1446439500896662016&$-0.0118$&$+24.5145$&$201.5676$&$4.7772$&$-1.6017$&$0.9760$&$19.6123$&$1.0080\text{E}\:\!\:\!\minus\:\!02$\\
88&3677279569234130048&$0.7083$&$-6.6177$&$193.6215$&$2.5629$&$-2.5554$&$0.8683$&$20.1710$&$1.0038\text{E}\:\!\:\!\minus\:\!02$\\
89&1443176738796291584&$0.2635$&$+23.2662$&$201.9803$&$4.6112$&$-0.9136$&$0.6071$&$17.8110$&$9.9017\text{E}\:\!\:\!\minus\:\!03$\\
90&3689436968086686592&$0.5275$&$+0.1543$&$195.5130$&$2.7430$&$-1.7675$&$0.6331$&$19.3850$&$9.8457\text{E}\:\!\:\!\minus\:\!03$\\
91&3743823932797933312&$-0.0844$&$+14.5741$&$199.2608$&$3.9796$&$-2.2242$&$0.8309$&$20.0393$&$9.6899\text{E}\:\!\:\!\minus\:\!03$\\
92&1458262618153176192&$0.2948$&$+32.7413$&$207.3543$&$5.1408$&$-0.3751$&$0.7559$&$18.9859$&$9.6086\text{E}\:\!\:\!\minus\:\!03$\\
93&3940420357056044288&$0.3519$&$+20.8439$&$199.4598$&$4.7599$&$-1.6054$&$0.6726$&$18.3198$&$8.9874\text{E}\:\!\:\!\minus\:\!03$\\
94&1507371072349542144&$-0.1597$&$+47.8610$&$215.9188$&$5.5026$&$-0.0514$&$0.8711$&$19.4371$&$8.8737\text{E}\:\!\:\!\minus\:\!03$\\
95&3938479650312918656&$0.6125$&$+17.5560$&$199.5950$&$4.4924$&$-1.4851$&$0.8348$&$19.5584$&$8.2065\text{E}\:\!\:\!\minus\:\!03$\\
96&1456251169001507584&$0.2697$&$+30.5236$&$204.2348$&$4.8161$&$-1.0784$&$0.6146$&$18.0593$&$7.8983\text{E}\:\!\:\!\minus\:\!03$\\
97&3938871179532201728&$0.1450$&$+18.9098$&$200.2058$&$4.2659$&$-0.6811$&$0.5948$&$18.5974$&$7.8864\text{E}\:\!\:\!\minus\:\!03$\\
98&3689019428546525440&$0.3076$&$-0.9438$&$195.4987$&$3.3426$&$-1.8699$&$0.6213$&$19.6326$&$7.8563\text{E}\:\!\:\!\minus\:\!03$\\
99&1633029475988441856&$0.0342$&$+65.8413$&$264.5104$&$1.8194$&$3.8462$&$0.8345$&$19.1037$&$7.6960\text{E}\:\!\:\!\minus\:\!03$\\
100&1635011792373656448&$0.1679$&$+65.1211$&$253.8823$&$2.7241$&$4.1788$&$0.5981$&$17.8074$&$7.5587\text{E}\:\!\:\!\minus\:\!03$\\
101&3744580156279785600&$0.2913$&$+16.0645$&$199.2184$&$3.8843$&$-1.3277$&$0.7673$&$20.0798$&$7.4507\text{E}\:\!\:\!\minus\:\!03$\\
102&3691972098303496192&$-0.2897$&$+1.9209$&$196.0260$&$3.1704$&$-2.0277$&$0.5907$&$18.0558$&$7.3961\text{E}\:\!\:\!\minus\:\!03$\\
103&3730783072057224064&$0.5447$&$+8.0236$&$196.8188$&$4.2337$&$-4.0880$&$0.6855$&$19.5750$&$6.9686\text{E}\:\!\:\!\minus\:\!03$\\
104&1443127570010678400&$0.3753$&$+22.8453$&$201.7464$&$4.9070$&$-0.6215$&$0.8291$&$19.4308$&$6.9297\text{E}\:\!\:\!\minus\:\!03$\\
105&3732743570009103488&$0.2201$&$+9.7921$&$197.2683$&$4.6305$&$-1.6454$&$0.7821$&$19.9412$&$6.7403\text{E}\:\!\:\!\minus\:\!03$\\
106&3736593170672106496&$0.5514$&$+11.7588$&$197.8533$&$5.0609$&$-3.1560$&$0.9104$&$19.8379$&$6.5168\text{E}\:\!\:\!\minus\:\!03$\\
107&1458436890741975936&$0.1259$&$+33.0617$&$206.2575$&$5.7670$&$-0.3490$&$0.8147$&$19.4588$&$6.4103\text{E}\:\!\:\!\minus\:\!03$\\
108&3685090289385421440&$-0.1285$&$-3.3308$&$195.3744$&$2.8122$&$-2.7455$&$0.6930$&$18.3146$&$6.3728\text{E}\:\!\:\!\minus\:\!03$\\
109&1456784049887562496&$-0.1976$&$+31.6569$&$205.4463$&$4.7322$&$-1.0968$&$0.8699$&$19.5363$&$6.2448\text{E}\:\!\:\!\minus\:\!03$\\
110&1505368857379439232&$0.1609$&$+43.9625$&$212.0110$&$6.5511$&$0.1754$&$0.9058$&$19.8964$&$6.1957\text{E}\:\!\:\!\minus\:\!03$\\
111&1496065511539578112&$0.1647$&$+38.4511$&$208.6889$&$5.5450$&$-0.0661$&$0.8831$&$19.9410$&$6.1851\text{E}\:\!\:\!\minus\:\!03$\\
112&1448703983748127104&$-0.1921$&$+27.2718$&$204.0384$&$5.1016$&$-1.2913$&$0.7961$&$19.1699$&$6.0476\text{E}\:\!\:\!\minus\:\!03$\\
113&3943652440204771328&$0.0945$&$+21.9936$&$199.5465$&$4.8426$&$-2.0650$&$0.6397$&$18.4953$&$5.8078\text{E}\:\!\:\!\minus\:\!03$\\
114&1628882805322936192&$0.1545$&$+63.7692$&$248.5402$&$3.4299$&$3.7673$&$0.6974$&$19.6397$&$5.7581\text{E}\:\!\:\!\minus\:\!03$\\
115&1605837320778662528&$0.3849$&$+52.3964$&$220.2742$&$4.9615$&$1.5494$&$0.8261$&$19.6347$&$5.6522\text{E}\:\!\:\!\minus\:\!03$\\

\bottomrule
\end{tabular}
\end{center}

\label{sel1}
\end{table*}


\bsp	
\label{lastpage}

\end{document}